\documentclass[
	twocolumn,
	prb,
	showpacs,
	preprintnumbers,
	superscriptaddress,
	amsmath,
	amssymb,
	citeautoscript
	]{revtex4-1}

\usepackage{amsmath}
\usepackage{graphicx}
\usepackage{epstopdf}
\usepackage{dcolumn}
\usepackage{color}
\usepackage{bm}

\newcommand{\dd}{{\mathbf d}}
\newcommand{\T}{\intercal}

\DeclareMathOperator{\sgn}{sgn}

\begin{document}

\title{Asymptotic behavior of impurity-induced bound states in  low-dimensional topological superconductors}

\author{V. Kaladzhyan}
\email{vardan.kaladzhyan@cea.fr}
\affiliation{Institut de Physique Th\'eorique, CEA/Saclay,
Orme des Merisiers, 91190 Gif-sur-Yvette Cedex, France}
\affiliation{Laboratoire de Physique des Solides, CNRS, Univ. Paris-Sud, Universit\'e Paris-Saclay, 91405 Orsay Cedex, France}
\author{C. Bena}
\affiliation{Institut de Physique Th\'eorique, CEA/Saclay,
Orme des Merisiers, 91190 Gif-sur-Yvette Cedex, France}
\affiliation{Laboratoire de Physique des Solides, CNRS, Univ. Paris-Sud, Universit\'e Paris-Saclay, 91405 Orsay Cedex, France}
\author{P. Simon}
\email{pascal.simon@u-psud.fr}
\affiliation{Laboratoire de Physique des Solides, CNRS, Univ. Paris-Sud, Universit\'e Paris-Saclay, 91405 Orsay Cedex, France}

\date{\today}

\begin{abstract}
We study theoretically the asymptotic behavior of the Shiba bound states associated with magnetic impurities embedded in both 2D and 1D anomalous superconductors. We calculate analytically the spatial dependence of the local density of states together with the spin polarization associated with the Shiba bound states. We show that the latter quantity exhibits drastic differences between s-wave and different types of p-wave superconductors. 
Such properties, which could be measured using spin-polarized STM, offer therefore  a way to discriminate between singlet and triplet pairing in low-dimensional superconductors, as well as a way  to estimate the amplitude of the triplet pairing in these systems.\\

\noindent{\it Keywords:} unconventional superconductor, Shiba states, localized impurity

\end{abstract}

\keywords{<Shiba states>}

\maketitle

\section{Introduction}
 Despite having been discovered more than thirty years ago \cite{Steglich1979,Jerome1980}, research on anomalous superconductors (SCs) remains a very active field  in condensed matter. This is in part related to the fact that some unconventional SCs such as strontium ruthenate \cite{Maeno2003} or doped topological insulators \cite{Sasaki2011} may offer a natural platform for topological superconductivity and therefore for Majorana fermions \cite{Qi2011,Leijnse2012,Ando2015}. Besides searching for intrinsically anomalous superconductors, a very promising alternative strategy consists in engineering topological superconductivity starting from traditional and well-characterized materials. Thus it has been proposed theoretically that arrays of  magnetic impurities or nanoparticles  on the surface of a conventional s-wave superconductor  may give rise to 1D and 2D topological superconductivity \cite{Choy2011,Nakosai2013,Yazdani2013,Braunecker2013,Klinovaja2013,Vazifeh2013,Pientka2013,Pientka2014,Poyhonen2014,Heimes2014,Reis2014,Brydon2015,Ojanen2015,Peng2015,Rontynen2015,Braunecker2015,Zhang2016,Hoffman2016}. Moreover, lattices of non-magnetic impurities on p-wave superconductors have also been proposed to realize p-wave superconductivity with a high Chern number \cite{Neupert2016,Kimme2016,Kaladzhyan2016b}. 
On the experimental side, zero bias peaks at the extremity of a chain of iron adatoms deposited on top of lead have recently been observed, consistent with  the predicted Majorana bound states \cite{Nadj-Perge2014,Pawlak2015,Ruby2015}.

The building block in the aforementioned strategy is the single impurity: a magnetic moment in a s-wave superconductor  gives rise to  so-called in-gap Shiba bound states (SBSs) \cite{Yu1965,Shiba1968,Rusinov1969,Balatsky2006} while SBSs can also emerge  around scalar impurities in p-wave superconductors \cite{Wang2004,Eremin2008,Nagai2014,Lutchyn2015,Kaladzhyan2015}. SBSs have been observed experimentally  using scanning tunneling microscopy (STM) \cite{Yazdani1997,Shuai-Hua2008,Franke2015}. It is worth noting that in these experiments the SBSs were found to be strongly localized around  the impurity (the spatial extent of the SBS wave function is of order $O(1\rm nm)$) 
However, in  recent experiments carried out in [\onlinecite{Menard2015}], a very large spatial extent of the Shiba wave function (of order O(20 nm)) was found for a magnetic impurity immersed in a  2D conventional SC. A partial explanation for this long-range extent is related to  the reduced dimensionality of the host superconductor. The  local density of states (LDOS) associated with the SBS decays as 
$1/r^2$  in a 3D SC, as $1/r$ in a 2D SC, and it does not depends on the distance $r$ from the impurity in a clean 1D superconductor (we have  typically in mind a proximitized wire). Such a slow decay makes the information stored in the decay length and in the period of the Friedel oscillations associated with the SBS more accessible experimentally. As we show in this paper, such information turns out to be useful to better characterize the host bulk superconductor particularly when the latter is an anomalous SC.

In a previous paper \cite{Kaladzhyan2015}, we have analyzed numerically the  behavior of the SBS in 2D p-wave  SCs. Here, we provide analytical expressions for the asymptotic behavior of the LDOS (both non-polarized and spin-polarized) associated with the SBSs induced 
 by single localized impurities (scalar or magnetic) not only in various 2D superconductors but also in 1D proximetized  superconductors. We give also the analytical form of the Shiba wave functions, which are essential for studying topological phases of matter engineered with adatom lattices and chains. They are for example used for computing the Chern numbers in  such emergent topological superconductors.\cite{Pientka2013,Poyhonen2014,Brydon2015,Ojanen2015} 
 
 More specifically, we focus mainly on superconductors with a triplet pairing (of p-wave type) which are characterized by the so-called $\dd$ vector, that determines the plane in which the spins of the Cooper pairs lie. We show that the spin-polarized local density of states (SP LDOS) allows not only to determine whether the host superconmductor has a dominant p-wave pairing, but also to discriminate between different directions of the  $\dd$ vector. This is particularly revealed in the Fourier transform of the SP LDOS of some Shiba states where the orbital nature of p-wave superconductivity naturally pops up.
In addition, our calculations show how the triplet pairing parameter alters both the period of the Friedel oscillations and the superconducting decay length scale, which may, in principle, allow to estimate the triplet pairing amplitude by carrying out an experiment similar to the one in [\onlinecite{Menard2015}]. We did not include the spin-orbit coupling in the present analysis.
A thorough analysis of its impact is provided in [\onlinecite{Kaladzhyan2015}] for a p-wave superconductor and in [\onlinecite{Kaladzhyan2016}] for a s-wave superconductor.

The paper is organized as follows: in Section II, we describe our general model and introduce the methods used in further calculations. In Section III, we  reconsider the case of 2D s-wave SC, although it has already been discussed e.g. in [\onlinecite{Brydon2015}]. This allows to fix our notations but also us to provide a comparative study noting that these previous results do not fully coincide with the ones presented below. In Section IV, we 
consider Shiba states in 2D superconductors with two different types of triplet pairing and analyze the asymptotic behavior
of both the non-polarized and SP LDOS.
 Similarly, Section V contains the description of SBSs in 1D superconductors with different types of order parameter. Finally, we provide a short conclusion in Section VI and leave some technical details to appendices.

\section{Model Hamiltonian} \label{General}

We consider either a 2D SC lying in the $(x,y)$ plane (see Sections \ref{2Dswave} and \ref{2Dpwave}) or a 1D superconducting wire directed along the $x$-axis (see Section \ref{1Dsystem}). The Hamiltonian for these two systems can be written in a general form in the Nambu basis  $\Psi_{\bm k}=(\psi_{\uparrow {\bm k}},\psi_{\downarrow {\bm k}},\psi^{\dag}_{\downarrow{-\bm k}},-\psi^{\dag}_{\uparrow {-\bm k}})^T$ as:
\begin{equation}
\mathcal{H}_0(\bm{k}) =	
	\begin{pmatrix}
		\xi_{\bm{k}} \sigma_0 & \Delta(\bm{k}) \\
		\Delta^\dag(\bm{k}) & -\xi_{\bm{k}} \sigma_0 \\
	\end{pmatrix},	
\label{H0}
\end{equation}
where 
\begin{equation}
\Delta(\bm{k}) = \Delta_s \sigma_0 + \varkappa\, \bm{d}(\bm{k}) \cdot \bm{\sigma},
\end{equation}
 is a general pairing function with a s-wave (singlet) component $\Delta_s$ and a p-wave (triplet) component $\varkappa$. Below we consider either the case of pure singlet pairing ($\varkappa=0$) or  pure triplet pairing ($\Delta_s=0$). The Pauli matrices ${\bm\sigma}$ are acting in the spin subspace, the operator $\psi^\dag_{\sigma \bm k}$ creates a particle of spin $\sigma=\uparrow,\downarrow$ of momentum ${\bm k}\equiv(k_x,k_y)$ for the 2D limit and ${\bm k}\equiv k$ in 1D.  The vector  $\dd(\bm{k})$  parametrizes the odd-parity triplet pairing term and will be discussed in detail in subsection \ref{2Dpwave}. Note that the concept of the $\dd$ vector is not usually introduced for 1D systems, but we  use it here  to simplify our notation. The energy dispersion in the normal state is given by $\xi_{\bm k} \equiv \frac{\bm{k}^2}{2m}-\varepsilon_F $.
This dispersion is a low-energy approximation for the tight binding Hamiltonian on the square lattice model that we used to obtain the numerical results in [\onlinecite{Kaladzhyan2015}]. 
We thus expect the analytical results in this paper to qualitatively reproduce  the numerical results in [\onlinecite{Kaladzhyan2015}] in the infrared limit.

Our goal is to study the effect of a single localized magnetic impurity on the system described above. Such impurity has both a scalar component $U$ and a magnetic component ${\bm J} = (J_x,J_y,J_z)$, and can be taken into account by means of the Hamiltonian:
\begin{equation}
\mathcal{H}_{imp} = V \delta(\bm{r}) \equiv
	\begin{pmatrix}
		U \sigma_0 + \bm{J} \cdot \bm{\sigma} & 0 \\
		0 & -U \sigma_0 + \bm{J} \cdot \bm{\sigma}
	\end{pmatrix} \delta(\bm{r}),
\label{Himp}
\end{equation}
where $U$ and ${\bm J}$ are the scalar and magnetic components respectively, $\bm r \equiv (x,y)$ in the 2D limit, and $\bm r \equiv x$ in 1D. The delta-like form of the impurity potential implies that the scattering occurs only in the $s$-channel. Also, in what follows we consider only classical impurities (e.g. we neglect quantum effects giving rise to phenomena such as the Kondo effect). 
In 2D the spin of the impurity can be decomposed without losing generality into an out-of-plane component  plus an in-plane component (in 1D the equivalent decomposition is into a component perpendicular to the wire and one along the wire). In what follows, we will therefore consider an impurity spin oriented either
along the z-axis, ${\bm J} = (0,0,J_z)$, or along the x-axis  ${\bm J} = (J_x,0,0)$. These two limits are generic enough to capture all the relevant physics.

In order to find the energy levels of the Shiba states 
we 
follow  the method introduced  in [\onlinecite{Pientka2013}] and seek the eigenvalues using:
\begin{equation}
	\left[\mathbb{I}_4-V G_0(E,{\bm r}={\bm 0}) \right]\Phi(\bm r=\bm 0) = 0.
\label{eigenv}
\end{equation}
The corresponding eigenfunctions are obtained by using
\begin{equation}
	\Phi (\bm r) = G_0(E, \bm r) V \Phi(\bm 0).
\label{eigenf}
\end{equation}
We note that to this purpose we need the explicit form of the Green's function at $\bm r = \bm 0$, as well as for $\bm r \neq \bm 0$. The form of the Green's function in real space can be obtained by a Fourier transform of the momentum space Green's function. The unperturbed retarded Green's function in momentum space can be written as $G_0(E,{\bm k})=\left[(E+i\delta)\mathbb{I}_4-\mathcal{H}_0({\bm k})\right]^{-1}$, where we have introduced a finite inverse quasiparticle lifetime $\delta$ (while this is kept finite in the numerical simulations \cite{Kaladzhyan2015}, it will be set to zero in the final results of the analytical calculations.) 

Once the eigenfunctions $\Phi (\bm r)$ are found, we can compute the full local density of states (LDOS), as well as the spin-polarized 
local density of states (SP LDOS) for the Shiba states using
\begin{equation}
	\rho(E, \bm r) = \Phi^\dag(\bm r) 
		\begin{pmatrix} 	
			0 & 0 \\ 
			0 & \sigma_0
		\end{pmatrix} \Phi (\bm r),
\label{LDOS}
\end{equation}
and
\begin{equation}
	\bm S(E, \bm r) = \Phi^\dag(\bm r) 
		\begin{pmatrix} 	
			0 & 0 \\ 
			0 & \bm \sigma
		\end{pmatrix} \Phi (\bm r),
\label{SPLDOS}
\end{equation}
where we take into account only the hole components of the spinor wave function. We focus only on these components. 
We note however that there is no qualitative difference between the hole components and electronic ones. The integration over the real-space coordinates in (\ref{LDOS}-\ref{SPLDOS}) gives access to the average total DOS and correspondingly to the average spin polarization of the Shiba states.

In the following sections we study the formation of Shiba states first in 2D superconducting materials and subsequently in 1D superconducting wires. 


\section{Shiba states in a 2D pure s-wave superconductor} \label{2Dswave}

We begin by considering the pure s-wave case, i.e. $\varkappa = 0$.
As mentioned above, this situation has already been addressed in previous works such as [\onlinecite{Brydon2015}] and [\onlinecite{Lutchyn2015}]. However, we are revisiting this limit here since the results presented in the previous references are not fully general, and contain as well inaccuracies that do not allow one to have a completely correct and general form for the non-polarized and SP LDOS of the Shiba states in such systems. 

In order to obtain the real space form of the retarded Green's functions we need to integrate the momentum space Green's function over all momenta. For this we need first to perform the following two integrals:
\begin{align}
X_0(\bm 0) &= -\int \frac{d\bm{k}}{(2\pi)^2} \frac{1}{\xi_{\bm k}^2+\omega^2}, \\
X_1(\bm 0) &= -\;\mathrm{p.v.}\negthickspace\int \frac{d\bm{k}}{(2\pi)^2} \frac{\xi_{\bm k}}{\xi_{\bm k}^2+\omega^2},
\end{align}
where $\omega^2 = \Delta_s^2 - E^2$. Using the principal value (abbreviated as $\mathrm{p.v.}$) for the second integral is fully equivalent to performing the calculation with a natural UV energy cut-off, such as the Debye frequency $\omega_D$, and then taking the limit of $\omega_D \to \infty$. We rewrite $\int \frac{d\bm{k}}{(2\pi)^2} = \nu \int d\xi_{\bm k}$, where $\nu = \frac{m}{2\pi}$, and we find
\begin{align}
X_0(\bm 0) = -\pi \nu \frac{1}{\omega}, \quad X_1(\bm 0) = 0.
\end{align}
Therefore, the bare Green's function is given by
\begin{align}
G_0(E,{\bm r}={\bm 0}) = -\frac{\pi\nu}{\omega}
		\begin{pmatrix} 	
				E \sigma_0 & \Delta_s \sigma_0 \\ 
				\Delta_s \sigma_0 & E \sigma_0
		\end{pmatrix}.
\end{align}
Using (\ref{eigenv}) it is easy to show that there are no sub-gap states for a purely scalar impurity $(\bm J = \bm 0)$; while in the case of a purely magnetic impurity $(U = 0)$ we obtain two energy levels independent of the direction of $\bm J$:
\begin{align}
	E_{1,\bar{1}} = \pm \frac{1-\alpha^2}{1+\alpha^2} \Delta_s, \; \text{where}\; \alpha = \pi \nu J.
\end{align}
The presence of two symmetric energy levels is a direct consequence of the imposed particle-hole symmetry of the Bogoliubov-de-Gennes Hamiltonian. 
The value $\alpha_c=1$ corresponds to a change in the ground state parity.

The corresponding eigenvectors are given by
\begin{align}
	\Phi_{\bar{1}}(\bm 0) = \begin{pmatrix} 1 & 0 & -1 & 0 \end{pmatrix}^\T, \; \Phi_1(\bm 0) = \begin{pmatrix} 0 & 1 & 0 & 1 \end{pmatrix}^\T
\end{align}
for an  impurity along the $z$-axis and
\begin{align}
	\Phi_{\bar{1}}(\bm 0) = \begin{pmatrix} 1 & 1 & -1 & -1 \end{pmatrix}^\T, \; \Phi_1(\bm 0) = \begin{pmatrix} 1 & -1 & 1 & -1 \end{pmatrix}^\T
\end{align}
for an  impurity along the $x$-axis. 

To find the coordinate dependence and the asymptotic behavior of the Shiba states we perform the Fourier transforms:
\begin{align}
	X_0(\bm r) &= -\int \frac{d\bm{k}}{(2\pi)^2} \frac{e^{i \bm {k r}}}{\xi_{\bm k}^2+\omega^2} ,\\
	X_1(\bm r) &= -\int \frac{d\bm{k}}{(2\pi)^2} \frac{\xi_{\bm k}\, e^{i \bm {k r}}}{\xi_{\bm k}^2+\omega^2}.
\end{align}
We detail this calculation in appendix A and here we only give the final results:
\begin{align}
	X_0(r) &= -2 \nu \cdot \frac{1}{\omega} \cdot \Im \left(K_0\left[-i(1+i\Omega)k_F r \right]\right),
	\label{X0s}
\end{align}
\begin{align}
	X_1(r) &= -2 \nu \cdot \Re \left(K_0\left[-i(1+i\Omega)k_F r \right]\right), \phantom{\cdot \frac{1}{\omega}!}
	\label{X1s}
\end{align}
where $\Omega \equiv \frac{\omega}{v_F k_F}$ and $K_0$ denotes the modified Bessel function of the second kind. It is worth noting that these functions diverge at $r=0$, but this divergence can be disregarded as it occurs only at the point where the impurity is localized and, therefore, the Schr\"odinger equation is not well-defined. However, this problem can be always avoided by introducing an infrared cut-off if needed. Since these functions have spherical symmetry we can write down the unperturbed Green's function as
\begin{equation*}
G_0(E,r) =	
	\begin{pmatrix}
		\left[E X_0(r) + X_1(r) \right] \sigma_0 & \Delta_s X_0(r) \sigma_0 \\
		\Delta_s X_0(r) \sigma_0 & \left[E X_0(r) - X_1(r) \right] \sigma_0 \\
	\end{pmatrix},
\end{equation*}
where $r = |\bm r|$. Using (\ref{eigenf}) we find for an impurity along the $z$-axis:
\begin{align*}
	\Phi_{\bar{1}}(r) &= +J_z \begin{pmatrix} (E_{\bar{1}}-\Delta_s)X_0(r)+X_1(r) \\ 0 \\ -(E_{\bar{1}}-\Delta_s)X_0(r)+X_1(r)\\ 0 \end{pmatrix}, \\
	\Phi_1(r) &= -J_z \begin{pmatrix} 0 \\ (E_1+\Delta_s)X_0(r)+X_1(r) \\ 0 \\ (E_1+\Delta_s)X_0(r)-X_1(r) \end{pmatrix}.
\end{align*}
The formation of Shiba states implies the breaking of Cooper pairs, and subsequently the coupling of the electrons to the spin of the impurity. Therefore there is no physical reason for the Shiba states to be polarized in any other direction than the direction of the impurity spin. Thus we expect intuitively that $S^x_{1,\bar{1}}(r) = S^y_{1,\bar{1}}(r) = 0$ for both $\Phi_{1}, \Phi_{\bar{1}}$ and this is indeed the case. Moreover we have
\begin{align*}
	S^z_{\bar{1}}(r) &= +\rho_{\bar{1}}(r) = +J_z^2\left[ (E_{\bar{1}}-\Delta_s)X_0(r)+X_1(r) \right]^2, \\
	S^z_1(r) &= -\rho_1(r) = -J_z^2\left[ (E_1+\Delta_s)X_0(r)+X_1(r) \right]^2.
\end{align*}
Similarly, for an impurity along the x-axis we have
\begin{align*}
	\Phi_{\bar{1}}(\bm r) &= +J_x \begin{pmatrix} (E_{\bar{1}}-\Delta_s)X_0(r)+X_1(r) \\ (E_{\bar{1}}-\Delta_s)X_0(r)+X_1(r) \\ -(E_{\bar{1}}-\Delta_s)X_0(r)+X_1(r) \\ -(E_{\bar{1}}-\Delta_s)X_0(r)+X_1(r) \end{pmatrix}, \\
	\Phi_1(\bm r) &= -J_x \begin{pmatrix} (E_1+\Delta_s)X_0(r)+X_1(r) \\ -(E_1+\Delta_s)X_0(r)-X_1(r) \\ (E_1+\Delta_s)X_0(r)-X_1(r) \\ -(E_1+\Delta_s)X_0(r)+X_1(r)\end{pmatrix}.
\end{align*}
For the same reasons as before, we have $S^y_{1,\bar{1}}(r) = S^z_{1,\bar{1}}(r) = 0$, and 
\begin{align*}
	S^x_{\bar{1}}(r) &= +\rho_{\bar{1}}(r) = +2J_x^2\left[ (E_{\bar{1}}-\Delta_s)X_0(r)+X_1(r) \right]^2, \\
	S^x_1(r) &= -\rho_1(r) = -2J_x^2\left[ (E_1+\Delta_s)X_0(r)+X_1(r) \right]^2. 
\end{align*}
Note that all the functions given above are not normalized. This choice is made for the sake of simplicity; moreover, since we are only interested in the form of the spatial dependence, the overall normalization constant is not relevant for our analysis.  

The asymptotic forms of the functions $X_0$ and $X_1$ for $r \to \infty$ are derived in the appendix B, and are given by
\begin{align*}
	X_0(r) &\sim  -\sqrt{2\pi}\, \nu \cdot \frac{1}{\omega} \frac{\sin \left(k_F r + \pi/4\right)}{\sqrt{k_F r}}e^{-k_S r}, \\
	X_1(r) &\sim  -\sqrt{2\pi}\, \nu \cdot \frac{\cos \left(k_F r + \pi/4\right)}{\sqrt{k_F r}}e^{-k_S r},
\end{align*}
where $k_S = \Omega k_F = \omega/v_F$ is the inverse superconducting decay length,  and the Friedel oscillations have a period corresponding to the Fermi momentum $k_F$. 

We should note that these results agree only qualitatively with previous studies of the Shiba states in 
2D s-wave superconductors (see e.g. [\onlinecite{Brydon2015}], [\onlinecite{Lutchyn2015}]). First of all, unlike the previous results expressed in terms of Bessel functions of the first kind and Struve functions\cite{Brydon2015,Lutchyn2015}, the form that we find for the Shiba states wavefunctions can be written in terms of modified Bessel functions of the second kind. The crucial difference between our results and the previous calculations is that the Bessel functions of the first kind and the Struve functions of complex arguments actually diverge for $r \to \infty$! Therefore, an expression containing these functions cannot correctly capture the full behavior of the wavefunctions of the 
SBS\footnote{The calculations performed in these references are performed using a technique involving differentiation under the integral sign, whereas the corresponding integrals are not uniformly convergent. This explains the divergence of the results for $r \to \infty$.}.
 On the contrary, the functions $X_0(r)$ and $X_1(r)$, given by (\ref{X0s}-\ref{X1s}), display a consistent behavior for large values of $r$, namely they go to zero when $r \to \infty$. Another minor difference between our results and the ones in [\onlinecite{Brydon2015}] and [\onlinecite{Lutchyn2015}] is a difference in the phase shift of the oscillating terms in the asymptotic expansions at large $r$. 

\section{Shiba states in a 2D pure p-wave superconductor} \label{2Dpwave}

We now exploit the model introduced in section \ref{General} to study a pure p-wave SC for which we take $\Delta_s = 0$. 
We only consider triplet superconductors which are gapped.
Similar to our previous numerical analysis \cite{Kaladzhyan2015}, we study here different $\dd$ vectors describing the triplet p-wave SCs \cite{Bauer2012}.
We focus on two different types of $\dd$ vectors which are generic enough to describe all 2D unconventional triplet gaped superconductors: an in-plane $\dd$ vector, $\dd_\parallel(\bm{k}) = (k_y, \,-k_x, \, 0)$, which corresponds to an unconventional time-reversal-invariant  SC; and an out-of plane $\dd$ vector, $\dd_\perp(\bm{k}) = (0, \,0, \, k_x+ik_y)$ which corresponds to a time-reversal symmetry-breaking SC. The latter model has been used to describe the properties of $\rm Sr_2RuO_4$ \cite{Maeno2003}. 

Note that for these two $\dd$ vectors, the system is characterized by two conserved quantities which can be written as
 \begin{align}
M^z_\parallel &= L_z + \sigma_z/2 ~~{\rm for}~\dd_\parallel, \\
M^z_\perp &= L_z - \tau_z/2~~{\rm for}~\dd_\perp,
\end{align}
correspondingly. Here $\tau_z$ is the Pauli matrix acting in the particle-hole subspace and $\bf L = r \times p$ is the orbital momentum operator.

\subsection{Energies of Shiba states and Shiba wavefunctions at $\bm r=\bm 0$}
The eigenvalues corresponding to the energies of the Shiba states, as well as the Shiba wavefunctions at $\bm r=\bm 0$ are independent of the $\dd$ vector choice, and can be found using the method introduced  in [\onlinecite{Pientka2013}]:
\begin{equation}
	\left[\mathbb{I}_4-V G_0(E,{\bm r}={\bm 0}) \right]\Phi(\bm r=\bm 0) = 0.
\label{eigenv1}
\end{equation}
Therefore, the first step is to calculate analytically $G_0(E,{\bm r}={\bm 0})$.
For this we note that the spectrum of $\mathcal{H}_0(\bm k)$ is given by $\mathcal{E}(\bm k) = \pm\sqrt{\xi^2_{\bm k}+\varkappa^2 \bm k^2}$, with a triplet gap parameter $\Delta_t \equiv \frac{\varkappa k_F}{\sqrt{1+\tilde{\varkappa}^2}}$, where $\tilde{\varkappa}\equiv \varkappa / v_F$. We need to perform the following integrals: 
\begin{align}
X_0(\bm 0) &= -\int \frac{d\bm k}{(2\pi^2} \frac{1}{\xi_{\bm k}^2+\varkappa^2 \bm{k}^2 - E^2}, \\
X_1(\bm 0) &= -\;\mathrm{p.v.}\negthickspace\int \frac{d\bm k}{(2\pi)^2} \frac{\xi_{\bm k}}{\xi_{\bm k}^2+\varkappa^2 \bm{k}^2 - E^2}, \\
X^\pm_2(\bm 0) &= \pm \negthickspace\int \frac{d\bm k}{(2\pi)^2} \frac{i \varkappa k_\pm}{\xi_{\bm k}^2+\varkappa^2 \bm{k}^2 - E^2},
\end{align}
where $k_\pm = k_x \pm i k_y$ and the symbol '$\mathrm{p.v.}$' corresponds to the principal value. The last integral is zero due to the angular part. The second integral has a UV divergence thus we need to use a natural cut-off which, in this particular case, is equivalent to computing the principal value of the integral. We linearize $\xi_p$ around the Fermi level, and using the spherical symmetry of the integrals we change variables $\xi_k \approx v_F (k-k_F), ~\int \frac{d\bm k}{(2\pi)^2} = \nu \int d\xi_k$, where $\nu = \frac{m}{2\pi}$, and finally we obtain:
\begin{align}
X_0(\bm 0) &= -\frac{\pi \nu}{\sqrt{1+\tilde{\varkappa}^2}} \frac{1}{\sqrt{\Delta_t^2-E^2}},\\
X_1(\bm 0) &= \frac{\pi \nu}{\sqrt{1+\tilde{\varkappa}^2}} \frac{\Delta_t}{\sqrt{\Delta_t^2-E^2}} \frac{\tilde{\varkappa}}{\sqrt{1+\tilde{\varkappa}^2}}, \\
X^\pm_2(\bm 0) &= 0.
\end{align}
The Green's function for $\bm r = \bm 0$ then takes the form:
\begin{align}
	G_0(E,\bm r = \bm 0) = -\frac{\pi\nu}{\sqrt{1+\tilde{\varkappa}^2}}  \times \phantom{aaaaaaaaaaaaaaaa} \\	\nonumber
	\times \frac{1}{\sqrt{\Delta_t^2-E^2}} \begin{pmatrix} 	
				\left(E-\gamma \Delta_t \right) \sigma_0 & 0 \\ 
				0 & \left(E+\gamma \Delta_t \right) \sigma_0
		\end{pmatrix}, 
\end{align}
where $\gamma \equiv \frac{\tilde{\varkappa}}{\sqrt{1+\tilde{\varkappa}^2}}$. Using this form for the Green's function and (\ref{eigenv1}) we compute below the eigenvalues and eigenfunctions for $\bm r = \bm 0$ for different types of impurities, as in Section \ref{2Dswave}.

\subsubsection{Scalar impurity}
 Unlike for pure s-wave SCs, in p-wave SCs a purely scalar impurity ($\bm J = \bm 0$) creates two pairs of degenerate Shiba states with energies 
\begin{align}
E_{\bar 1, \bar 2} &= - \frac{-\gamma \beta^2 + \sqrt{1+\beta^2 (1-\gamma^2)}}{1+\beta^2}\Delta_t, \\ 
E_{1,2} &= + \frac{-\gamma \beta^2 + \sqrt{1+\beta^2 (1-\gamma^2)}}{1+\beta^2}\Delta_t,
\end{align}
where $\beta = \frac{\pi \nu U}{\sqrt{1+\tilde{\varkappa}^2}}$, and eigenfunctions 
\begin{align}
	\Phi_{\bar 1}(\bm 0)=\begin{pmatrix} 0 & 1 & 0 & 0 \end{pmatrix}^\T, \;
	\Phi_{\bar 2}(\bm 0)=\begin{pmatrix} 1 & 0 & 0 & 0 \end{pmatrix}^\T, \\
	\Phi_{2}(\bm 0)=\begin{pmatrix} 0 & 0 & 0 & 1 \end{pmatrix}^\T, \;
	\Phi_{1}(\bm 0)=\begin{pmatrix} 0 & 0 & 1 & 0 \end{pmatrix}^\T.
\end{align}

A possible explanation of the existence of these states is that a p-wave SC contains Cooper pairs with non-zero angular momentum due to the triplet pairing, and thus there are intrinsic magnetic fields impossible to observe unless one introduces a defect into the system, e.g. an impurity of any type. 
While in the case of a p-wave SC with a non-magnetic impurity we seem to have two pairs of degenerate states, we can think about this situation as having only two Shiba bound states within the gap mixing particle and hole degrees of freedom.
Because the particle and hole components are the parts
of the same state, they appear symmetrically in energy relative to the
chemical potential, the positive and negative energy counterparts corresponding to the particle and hole component of the same bound state wave function respectively \cite{Sakurai1970,Schrieffer1997,Balatsky2006,Aguado2015}.

\subsubsection{Magnetic impurity} 
Since in the case of a purely magnetic impurity ($U=0$) two types of coupling between the Cooper pairs and the impurity are possible, there are four Shiba states with energies independent of the impurity spin direction:
\begin{align}
E_{1,\bar 1} &= \pm \frac{\gamma \alpha^2 + \sqrt{1+\alpha^2 (1-\gamma^2)}}{1+\alpha^2}\Delta_t, \\
E_{2,\bar 2} &= \pm \frac{-\gamma \alpha^2 + \sqrt{1+\alpha^2 (1-\gamma^2)}}{1+\alpha^2}\Delta_t,
\end{align}
where $\alpha = \frac{\pi \nu J}{\sqrt{1+\tilde{\varkappa}^2}}$. For weak impurities these levels are ordered as follows $E_{\bar 1} < E_{\bar 2} < E_{2} < E_{1}$, while for a stronger impurities the middle levels exchange places, changing the order to $E_{\bar 1} < E_{2} < E_{\bar 2} < E_{1}$. 

\begin{figure}
	\includegraphics*[width=\columnwidth]{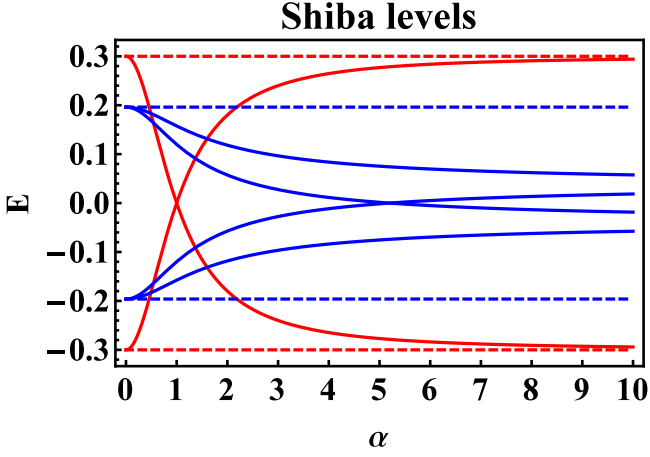}
	\caption{The energies of the Shiba states (in arbitrary units) for a s-wave SC (red lines) and a p-wave SC (blue lines) as function of the dimensionless impurity strength $\alpha=\pi\nu J$. The red and blue dashed lines denote the s-wave and p-wave SC gaps correspondingly. We set $k_F=1, \, \Delta_s = 0.3, \, \tilde{\varkappa}=0.2$.}
	\label{shibalevels}
\end{figure}

The behaviour of these energy levels is qualitatively different in s-wave SCs than in p-wave SCs. First of all, when increasing the impurity strength, the Shiba states in a s-wave SC approach the gap and eventually merge with the continuum, whereas in the p-wave case they remain in the gap and asymptotically approach $\pm \gamma \Delta_t $ (see Fig. ~\ref{shibalevels}). Second, the crossing point in the s-wave case is always at $\alpha=1$ independent of the singlet pairing $\Delta_s$, while for p-wave SCs the crossing point appears at $\alpha = 1/\gamma \gg 1$ and thus depends on the value of the triplet pairing $\tilde{\varkappa}$. Some realistic values of $\alpha$ can be extracted from experimental data given e.g. in [\onlinecite{Menard2015}] for an s-wave SC: the superconducting gap is about $1\,\mathrm{meV}$ and the Shiba state appears at $0.1\,\mathrm{meV}$, therefore $\alpha \approx 0.9$ (close to the crossing point in figure~\ref{shibalevels}). Since no p-wave superconductor has been unambiguously discovered so far (there are only some candidates like Sr$_2$RuO$_4$\cite{Maeno2003}), there is no experimental data available. However, taking comparable impurity strengths, we therefore expect the experiments to be in the regime much before the crossing point (see figure~\ref{shibalevels}). 
We believe that it is unlikely to observe that point  experimentally, because the dimensionless impurity strength must be too large ($\alpha \sim 10$ since $\gamma \ll 1$). Furthermore, in this regime the gap is renormalised (or even utterly suppressed), and the problem requires a self-consistent approach leading to a qualitatively different result, namely, the Shiba states might transform into the Andreev bound states (see [\onlinecite{Meng2015}] for further details). Also note that the physical meaning of the crossing point is the change in the ground state parity for both types of pairing.

For an impurity with spin along z-axis we have:
\begin{align}
	\Phi_{\bar 1}(\bm 0)=\begin{pmatrix} 0 & 0 & 1 & 0 \end{pmatrix}^\T, \;
	\Phi_{\bar 2}(\bm 0)=\begin{pmatrix} 1 & 0 & 0 & 0 \end{pmatrix}^\T, \\
	\Phi_{2}(\bm 0)=\begin{pmatrix} 0 & 0 & 0 & 1 \end{pmatrix}^\T, \;
	\Phi_{1}(\bm 0)=\begin{pmatrix} 0 & 1 & 0 & 0 \end{pmatrix}^\T.
\end{align}
For an impurity with spin along x-axis, we have:
\begin{align}
	\Phi_{\bar 1}(\bm 0)&=\begin{pmatrix} 0 & 0 & 1 & 1 \end{pmatrix}^\T, \;
	\Phi_{\bar 2}(\bm 0)=\begin{pmatrix} 1 & 1 & 0 & 0 \end{pmatrix}^\T, \\
	\Phi_2(\bm 0)&=\begin{pmatrix} 0 & 0 & 1 & -1 \end{pmatrix}^\T, \;
	\Phi_1(\bm 0)=\begin{pmatrix} 1 & -1 & 0 & 0 \end{pmatrix}^\T.
\end{align}

\subsection{Coordinate dependence of the Shiba wavefunctions} 
To find the spatial dependence of the Shiba states wavefunctions we 
use
\begin{equation}
	\Phi (\bm r) = G_0(E, \bm r) V \Phi(\bm 0).
\label{eigenf1}
\end{equation}
While $G_0(E, \bm r=\bm 0)$ is independent of the choice of  $\dd$, $G_0(E, \bm r\ne \bm 0)$, and thus the spatial dependence of the eigenfunctions, as well as the spatial dependence of the LDOS and SP LDOS change drastically with the choice of  $\dd$. 
In what follows for every choice of the $\dd$ vector we construct the retarded Green's function and the corresponding eigenfunctions for different types of impurities, for which we also compute all the polarized and non-polarized components of LDOS. However, we note first that for both choices of $\dd$ vector we need to perform the following integrations:
\begin{align}
X_0(\bm r) &= -\int \frac{d\bm k}{(2\pi)^2} \frac{e^{i \bm{k r}}}{\xi_{\bm k}^2+\varkappa^2 \bm{k}^2 - E^2}, \\
X_1(\bm r) &= -\int \frac{d\bm k}{(2\pi)^2} \frac{\xi_{\bm k}\, e^{i \bm{k r}}}{\xi_{\bm k}^2+\varkappa^2 \bm{k}^2 - E^2}, \\
X^\pm_2(\bm r) &= \pm \negthickspace\int \frac{d\bm k}{(2\pi)^2} \frac{i \varkappa k_\pm\, e^{i \bm{k r}}}{\xi_{\bm k}^2+\varkappa^2 \bm{k}^2 - E^2},
\end{align}
Below we give the results of the calculations, the details of which can be found in appendix A.
\begin{align}
X_0(r) &= -\frac{2\nu}{1+\tilde{\varkappa}^2} \cdot \frac{1}{\omega} \cdot \Im K_0 \left[ -i (1-\gamma^2+i\Omega) k_F r \right] \\
X_1(r) &= -\frac{2\nu}{1+\tilde{\varkappa}^2} \cdot \Im \left\{ \left(i -\frac{\gamma^2}{\Omega} \right) \times \right. \\ \nonumber
 & \phantom{aaaaaaa}\times \left. K_0 \left[ -i (1-\gamma^2+i\Omega) k_F r \right] \right\} \\
X^\pm_2(\bm r) &= \pm \frac{2\nu}{1+\tilde{\varkappa}^2} \cdot \frac{\varkappa k_F}{\omega} \cdot e^{\pm i \varphi_{\bm r}} \times \\ \nonumber
&\times \Re \left\{\left(1-\gamma^2+i\Omega\right) K_1 \left[ -i (1-\gamma^2+i\Omega) k_F r \right]\right\},
\end{align}
where we denote $\Omega \equiv \frac{\omega}{v_F k_F} = \frac{1}{v_F k_F} \frac{\sqrt{\Delta_t^2-E^2}}{\sqrt{1+\tilde{\varkappa}^2}}$, and 
$$
e^{\pm i \varphi_r} \equiv \frac{x \pm i y}{\sqrt{x^2+y^2}} = \frac{x \pm i y}{r}
$$
reflects all the characteristic asymmetry originating from the p-wave pairing orbital nature. We use the fact that $\Omega \ll 1$, which holds for all sub-gap energies. We give also the asymptotic behavior  of these integrals (see appendix B for a full derivation):
\begin{align}
	X_0(r) &\sim  - \frac{\sqrt{2\pi}\,\nu}{\sqrt{1+\tilde{\varkappa}^2}} \cdot \frac{1}{\sqrt{\Delta_t^2-E^2}} \frac{\sin \left(k'_F r + \pi/4\right)}{\sqrt{k'_F r}}e^{-k_S r}, \label{x1}\\
	X_1(r) &\sim  +\frac{\sqrt{2\pi}\,\nu}{1+\tilde{\varkappa}^2} \cdot \tilde{\varkappa} \frac{\Delta_t}{\sqrt{\Delta_t^2-E^2}} \cdot\frac{\sin \left(k'_F r + \pi/4\right)}{\sqrt{k'_F r}}e^{-k_S r},\label{x2} \\
	X_2^\pm(\bm r) &\sim \pm \frac{\sqrt{2\pi}\,\nu}{1+\tilde{\varkappa}^2} \cdot \frac{\Delta_t}{\sqrt{\Delta_t^2-E^2}} \cdot e^{\pm i \varphi_r} \times \label{x3}\\ \nonumber
	&\phantom{aaaaaaaaaaaaaaaaaaaa}\times \frac{\cos \left(k'_F r + \pi/4\right)}{\sqrt{k'_F r}}e^{-k_S r} ,
\end{align}
where $k_S = \Omega k_F = \frac{\sqrt{\Delta_t^2-E^2}}{v_F\sqrt{1+\tilde{\varkappa}^2}}$ is the inverse superconducting decay length scale, and $k'_F = \frac{k_F}{1+\tilde{\varkappa}^2}$. 
\vspace{.2in}

\subsubsection{ In-plane $\bm{d_\parallel}$} 
\vspace{.2in}
The retarded Green's function in this case can be written using the integrals given above:
\begin{equation*}
G_0(E,\bm r) =
	\begin{pmatrix}
		\left[E X_0(r) + X_1(r)\right]\sigma_0 & D_\parallel(\bm r)  \\
		D_\parallel(\bm r) & \left[E X_0(r) - X_1(r)\right]\sigma_0
	\end{pmatrix}, 
\end{equation*}
where we denote:
\begin{align*}
D_\parallel(\bm r) \equiv \begin{pmatrix} 0 & X^-_2(\bm r) \\ X^+_2(\bm r) & 0 \end{pmatrix}.
\end{align*}

The wavefunctions for the SBS arising for different types of impurities can be calculated subsequently using (\ref{eigenf1}).

{\it Scalar impurity.} In this case we find
\begin{align*} 
\Phi_{\bar 1}(\bm r) &= +U\begin{pmatrix} 0 \\ E_{\bar 1, \bar 2} X_0(r) + X_1(r) \\ X^-_2(\bm r) \\ 0 \end{pmatrix}, \\
\Phi_{\bar 2}(\bm r) &= +U\begin{pmatrix} E_{\bar 1, \bar 2} X_0(r) + X_1(r) \\ 0 \\ 0 \\ X^+_2(\bm r) \end{pmatrix}, \\ 
\Phi_{2}(\bm r) &= -U\begin{pmatrix} X^-_2(\bm r) \\ 0 \\ 0 \\ E_{1,2} X_0(r) - X_1(r) \end{pmatrix} \\
\Phi_{1}(\bm r) &= -U\begin{pmatrix} 0 \\ X^+_2(\bm r) \\ E_{1,2} X_0(r) - X_1(r) \\ 0 \end{pmatrix}.
\end{align*}
It is worth noting that the Hamiltonian in (\ref{H0}) with a scalar impurity described by (\ref{Himp}) with $\bf J=0$ still commutes with $M^z_\parallel$ and therefore the states described above are also the eigenstates of this operator, namely: $M^z_\parallel \Phi_{1, \bar 1} = +\frac{1}{2} \Phi_{1, \bar 1}$ and $M^z_\parallel \Phi_{2, \bar 2} = -\frac{1}{2} \Phi_{2, \bar 2}$. Therefore, we expect no explicit symmetry breaking nor any explicit p-wave orbital features to be observed in the full LDOS or in the SP LDOS. Indeed, we find that for all the states we have $S^x(\bm r) = S^y(\bm r) = 0$. Also, we note that the $z$-component of the SP LDOS and the LDOS are radially symmetric:
\begin{align*}
	S^z_{\bar 1}(r) &= +\rho_{\bar 1}(r) = -U^2 X^-_2(\bm r) X^+_2(\bm r) \geqslant 0, \\
%
	S^z_{\bar 2}(r) &= -\rho_{\bar 2}(r) = +U^2 X^-_2(\bm r) X^+_2(\bm r) \leqslant 0, \\
%
	S^z_{2}(r) &= -\rho_{2}(r) = -U^2\left[E_{1, 2} X_0(r) - X_1(r)\right]^2 \leqslant 0, \\
%
	S^z_{1}(r) &= +\rho_{1}(r) = +U^2\left[E_{1, 2} X_0(r) - X_1(r)\right]^2 \geqslant 0.
\end{align*}
We can see that the degenerate states have exactly opposite spin, and thus the total SP LDOS corresponding to the SBS energies, which is obtained by summing up over the two states with the same energy, is exactly zero, consistent also with the numerical simulations.

Moreover, when comparing the asymptotic behavior for the SP LDOS, as derived from the asymptotic expressions in (\ref{x1},\ref{x2},\ref{x3}), with the one obtained for the pure s-wave SC, we see that we have an additional factor $k'_F = \frac{k_F}{1+\tilde{\varkappa}^2}$ depending on the p-wave parameter $\varkappa$ that renormalizes the Fermi momentum and also changes the decay length scale. Such renormalization, if detected, may serve to measure the triplet pairing parameter by analysing the spatial structure of the SBS using STM.

{\it Magnetic impurity with spin $\parallel$ $z$.} For this type of impurity we find
\begin{align*}
\Phi_{\bar 1}(\bm r) = +J_z\begin{pmatrix} 0 \\ X^+_2(\bm r) \\E_{\bar 1} X_0(r) - X_1(r) \\ 0 \end{pmatrix}, \\
\Phi_{\bar 2}(\bm r) = +J_z\begin{pmatrix} E_{\bar 2} X_0(r) + X_1(r) \\ 0 \\ 0 \\ X^+_2(\bm r) \end{pmatrix}, \\
\Phi_2(\bm r) = -J_z\begin{pmatrix} X^-_2(\bm r) \\ 0 \\ 0 \\ E_2 X_0(r) - X_1(r) \end{pmatrix}, \\
\Phi_1(\bm r) = -J_z\begin{pmatrix} 0 \\ E_1 X_0(r) + X_1(r) \\ X^-_2(\bm r) \\ 0 \end{pmatrix}.
\end{align*}
Like in the case of a scalar impurity, we note that the Hamiltonian still commutes with $M^z_\parallel$, and therefore the states found above are also the eigenstates of $M^z_\parallel$, such that $M^z_\parallel \Phi_{\bar 1, \bar 2} = +\frac{1}{2} \Phi_{\bar 1, \bar 2}$, and $M^z_\parallel \Phi_{1,2} = -\frac{1}{2} \Phi_{1,2}$. For all the states $S^x(\bm r) = S^y(\bm r) = 0$. Below we give the expressions for the $z$-component  of the SP LDOS, and for the non-polarized LDOS, which are fully radially symmetric, same as for a scalar impurity:
\begin{align*}
	S^z_{\bar 1}(r) &= +\rho_{\bar 1}(r) = +J_z^2\left(E_{\bar 1} X_0(r) - X_1(r)\right)^2 \geqslant 0,\\
%
	S^z_{\bar 2}(r) &= -\rho_{\bar 2}(r) = + J_z^2 X^-_2(\bm r) X^+_2(\bm r) \leqslant 0, \\
%
	S^z_2(r) &= -\rho_2(r) = -J_z^2\left(E_2 X_0(r) + X_1(r)\right)^2 \leqslant 0, \\
%
	S^z_1(r) &= +\rho_1(r) = -J_z^2 X^-_2(\bm r) X^+_2(\bm r) \geqslant 0.
\end{align*}
We can see from this expressions that the average SPDOS, obtained by integrating these expressions over all space, is positive for the first and fourth states, and negative for the second and third. Thus, the analytical results are perfectly consistent with the numerical simulations given in [\onlinecite{Kaladzhyan2015}].

\begin{center}
\begin{figure}
	\includegraphics*[scale=0.5]{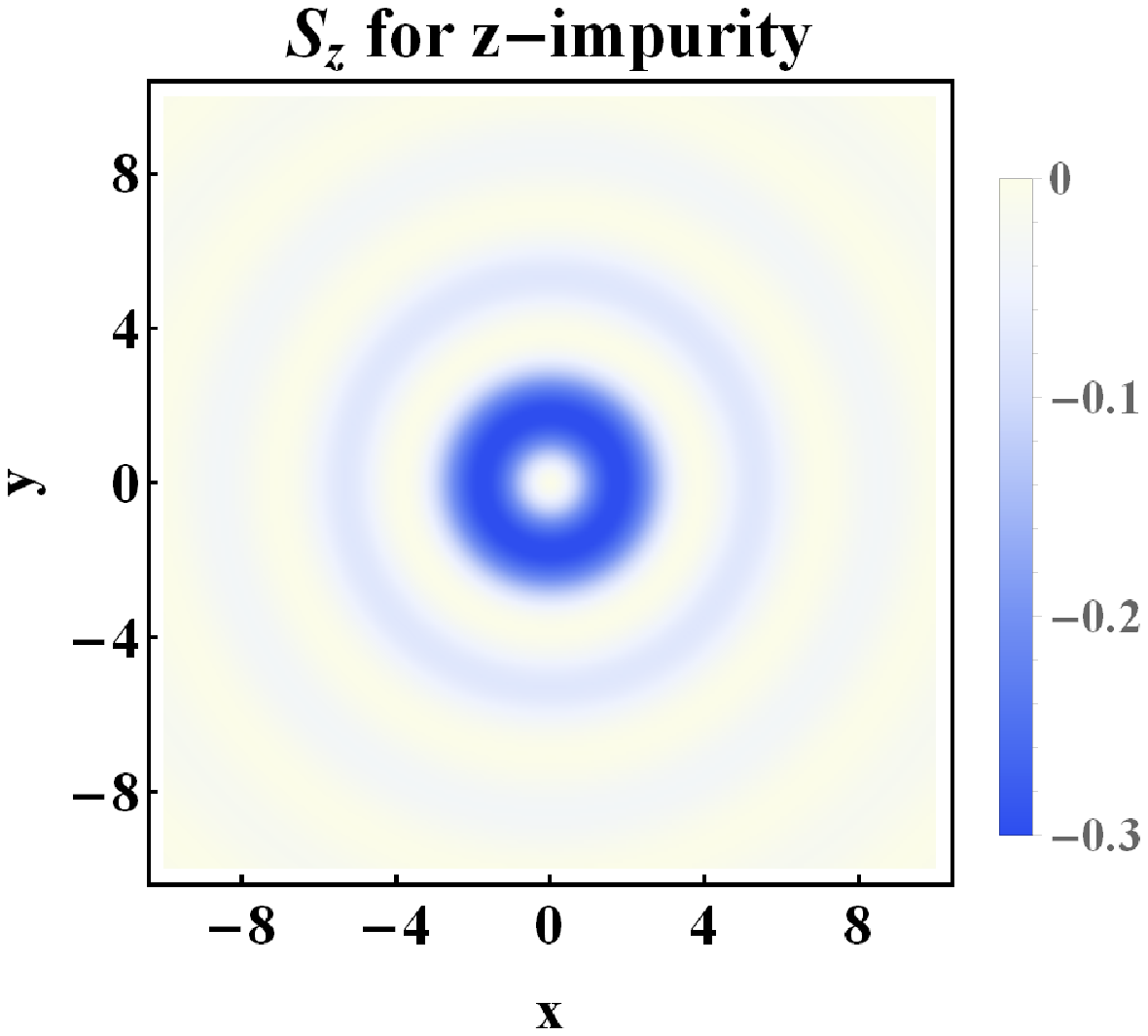} 
	\includegraphics*[scale=0.5]{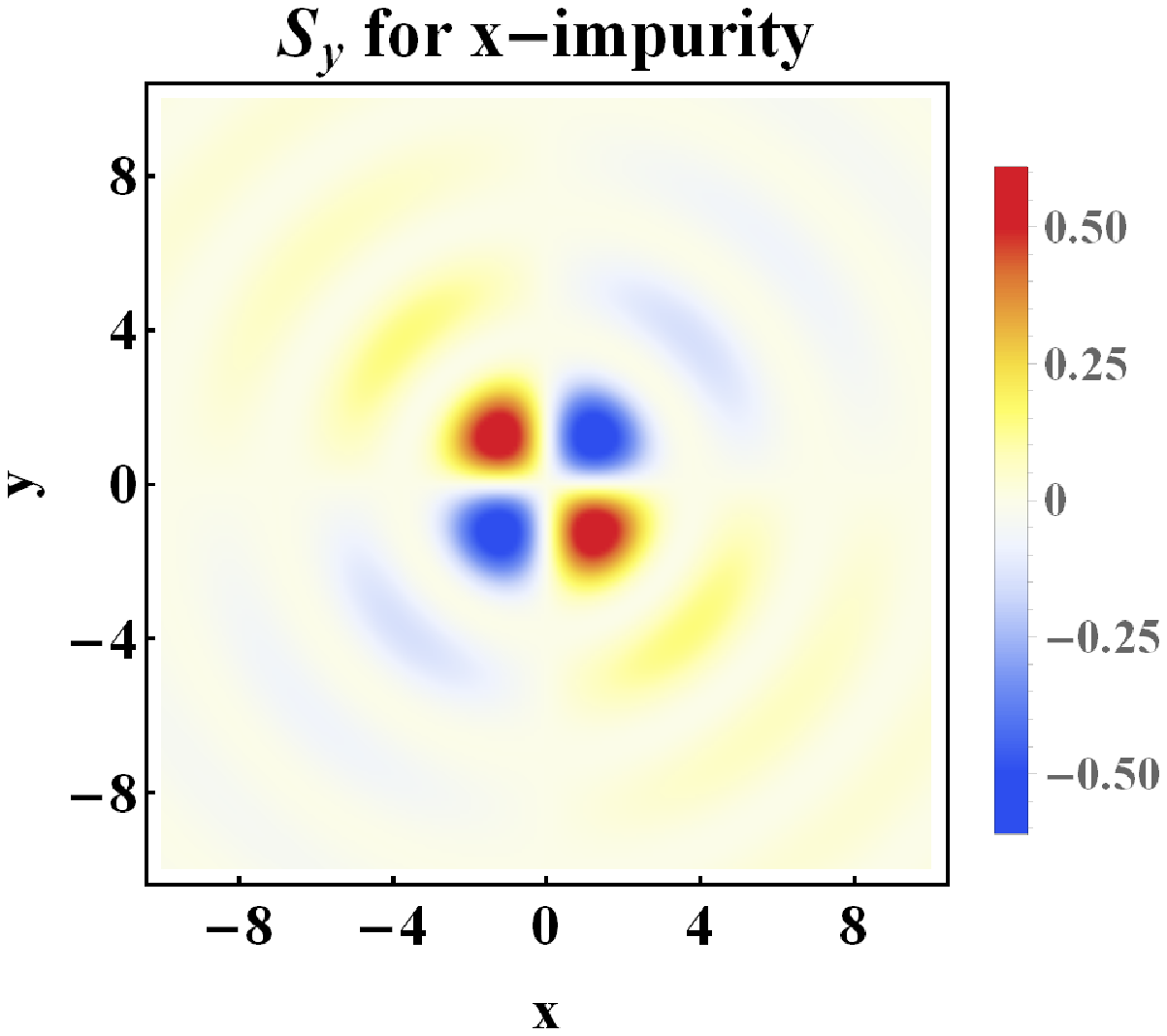}
	\caption{SP LDOS (in arbitrary units) in coordinate space, for an in-plane $\dd$ vector, and for an energy $E = E_{\bar 2}$. We consider a magnetic impurity with spin along $z$ (upper panel) and along $y$ (lower panel), with impurity strengths of $J_z = 2$ and $J_x = 2$ respectively. The SP LDOS in the upper panel is radially symmetric, whereas in the lower one it reflects the characteristic p-wave four-fold symmetry. We set $\Delta_s=0$, $\varkappa = 0.2$ and an inverse quasiparticle lifetime $\delta=0.01$.}
	\label{figure1}
\end{figure}
\end{center}

{\it Magnetic impurity with spin $\parallel$ $x$.}
Unlike the cases of a scalar impurity and of a magnetic impurity along $z$, the Hamiltonian describing a magnetic impurity with the spin along $x$ no longer commutes with $M^z_\parallel$ and therefore the SBS are not the eigenstates of this operator, and are thus expected to break the rotational symmetry that we have observed in the previous limits. Indeed we obtain:
\begin{align*}
\Phi_{\bar 1}(\bm r) = J_x\begin{pmatrix} X^-_2(\bm r) \\ X^+_2(\bm r) \\ +E_{\bar 1} X_0(r) - X_1(r) \\ +E_{\bar 1} X_0(r) - X_1(r) \end{pmatrix}, \\
\Phi_{\bar 2}(\bm r) = J_x\begin{pmatrix} +E_{\bar 2} X_0(r) + X_1(r) \\ +E_{\bar 2} X_0(r) + X_1(r) \\ X^-_2(\bm r) \\ X^+_2(\bm r) \end{pmatrix}, \\
\Phi_2(\bm r) = J_x\begin{pmatrix} X^-_2(\bm r) \\ -X^+_2(\bm r) \\ -E_2 X_0(r) + X_1(r) \\ +E_2 X_0(r) - X_1(r) \end{pmatrix}, \\
\Phi_1(\bm r) = J_x\begin{pmatrix} -E_1 X_0(r) - X_1(r) \\ + E_1 X_0(r) + X_1(r) \\ X^-_2(\bm r) \\ -X^+_2(\bm r) \end{pmatrix}.
\end{align*}
We exploit once more (\ref{LDOS}-\ref{SPLDOS}) to compute the LDOS and the SP LDOS and we find:
\begin{align*}
	S^x_{\bar 1}(r) &= +\rho_{\bar 1}(r) = +2J_x^2\left(E_{\bar 1} X_0(r) - X_1(r)\right)^2 \geqslant 0, \\
	S^y_{\bar 1}(\bm r) &= S^z_{\bar 1}(\bm r) = 0, \\
%
	S^x_{\bar 2}(\bm r) &= -J_x^2\left\{ \left[ X^+_2(\bm r) \right]^2 + \left[ X^-_2(\bm r) \right]^2 \right\},\\
	S^y_{\bar 2}(\bm r) &= +i J_x^2\left\{ \left[ X^+_2(\bm r) \right]^2 - \left[ X^-_2(\bm r) \right]^2 \right\}, \\
	S^z_{\bar 2}(\bm r) &= 0, \\
	\rho_{\bar 2}(r) &= -2J_x^2 X^-_2(\bm r) X^+_2(\bm r),\\
	S^x_2(r) &= -\rho_2(r) = -2J_x^2\left(E_2 X_0(r) - X_1(r)\right)^2 \leqslant 0, \\
	S^y_2(\bm r) &= S^z_2(\bm r) = 0,\\
%
	S^x_1(\bm r) &= +J_x^2\left\{ \left[ X^+_2(\bm r) \right]^2 + \left[ X^-_2(\bm r) \right]^2 \right\}, \\  	
	S^y_1(\bm r) &= -iJ_x^2\left\{ \left[ X^+_2(\bm r) \right]^2 - \left[ X^-_2(\bm r) \right]^2 \right\},\\ 	
	S^z_1(\bm r) &= 0, \\
	\rho_1(r) &= -2J_x^2 X^-_2(\bm r) X^+_2(\bm r).
\end{align*}
Indeed, we see that the $x$-components of the spin of the states $\bar{1}$ and $2$ are opposite in sign, while the rotational symmetry for these states is preserved.  However the states $\bar{2}$ and $1$ show peculiar orbital features characteristic for the p-wave, that we show in figure \ref{figure1} by plotting the corresponding SP LDOS. The rings of high intensity appearing in these figures correspond to Friedel oscillations with the wavevector $2k_F'$ defined above. 
The strong radially asymmetric behavior of the $S_x$ component for an $x$-impurity is consistent with the $\cos 2\phi_{\bm r}$ dependence arising in the asymptotic expansion of $\left[ X^+_2(\bm r) \right]^2 + \left[ X^-_2(\bm r) \right]^2$. 

Let us focus on the states $\bar{2}$ and $1$ and particularly on their average spin. 
Noticing that
$$
X^\pm_2(\bm r) = \pm e^{\pm i \varphi_r} F(r),
$$
where $F(r)$ has no angular dependence, we thus find
$$
\int d\bm r \left[ X^\pm_2(\bm r) \right]^2 = \int\limits_0^{+\infty} r F^2(r)dr \int\limits_0^{2\pi} e^{\pm 2 i \varphi_r} d\varphi_r = 0
$$
due to the angular part. Therefore, we find that the average spin for the states $\bar{2}$ and $1$ is exactly zero which is consistent with previous  numerical analysis \cite{Kaladzhyan2015}. This result can be directly traced back to the p-wave nature of the host superconductor which manifests in some of the Shiba states.

\begin{figure}
	\centering
	\begin{tabular}{cc}
		\includegraphics*[width=0.49\columnwidth]{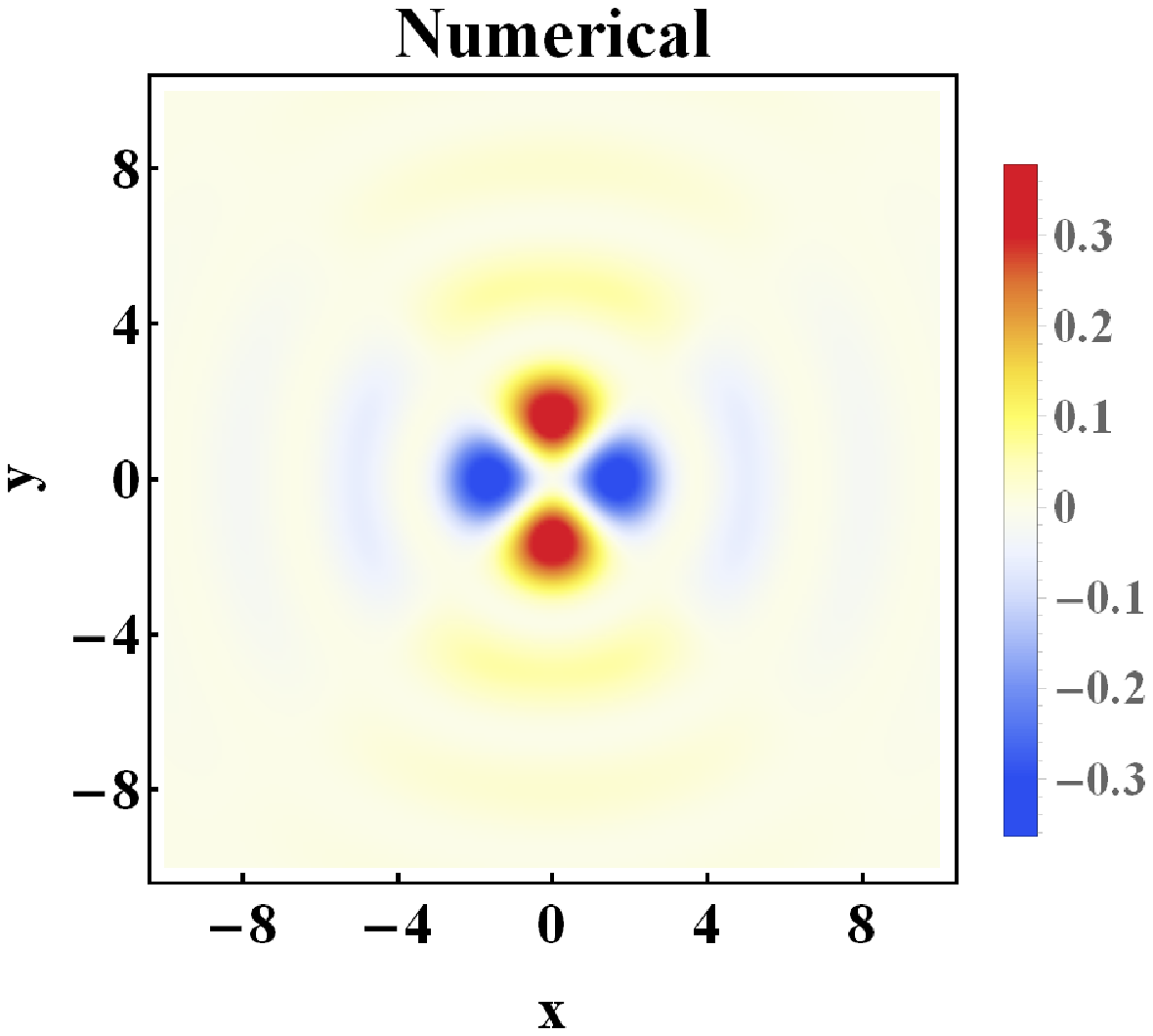}  & 											\includegraphics*[width=0.49\columnwidth]{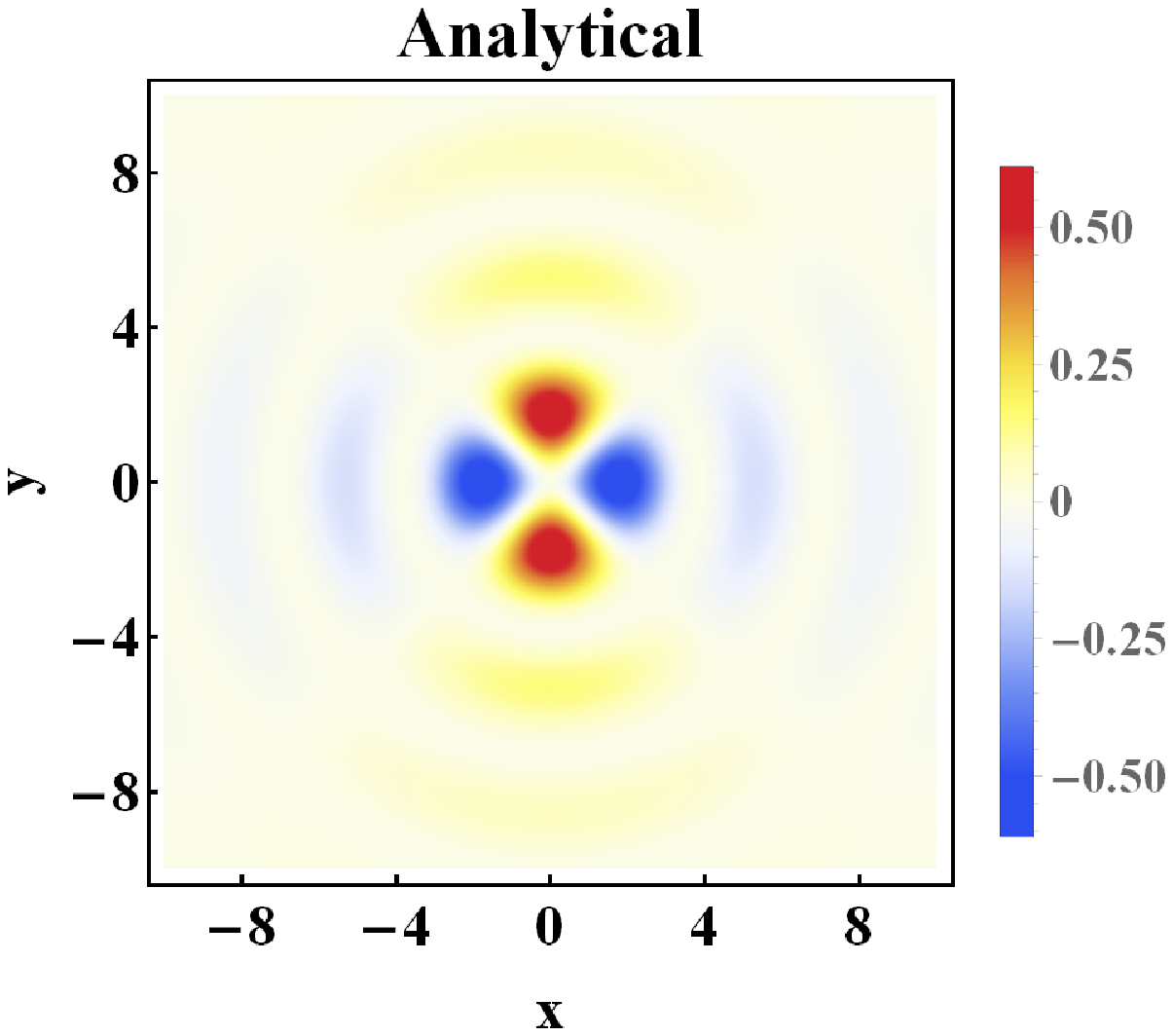} \\
		\includegraphics*[width=0.49\columnwidth]{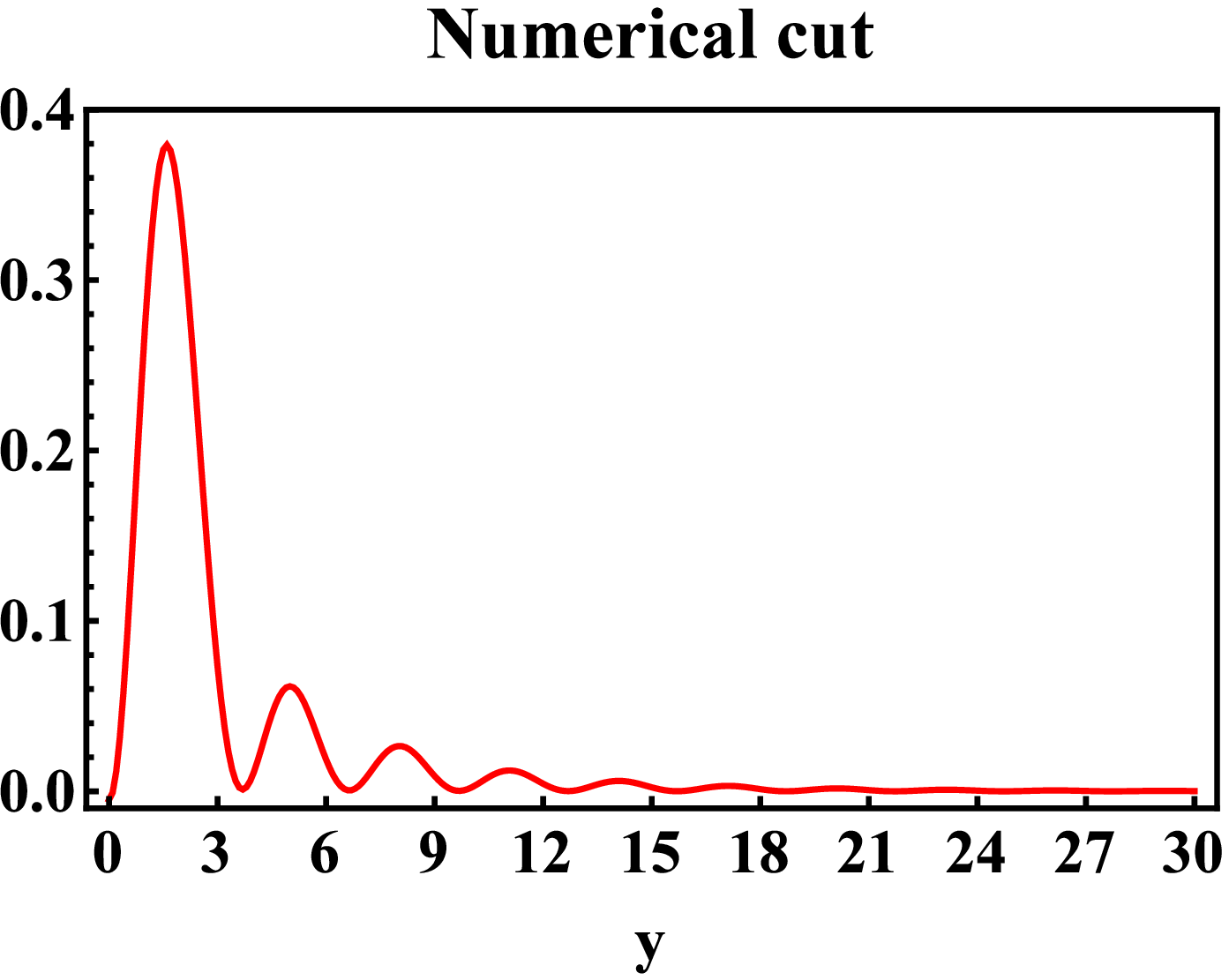}  & 											\includegraphics*[width=0.49\columnwidth]{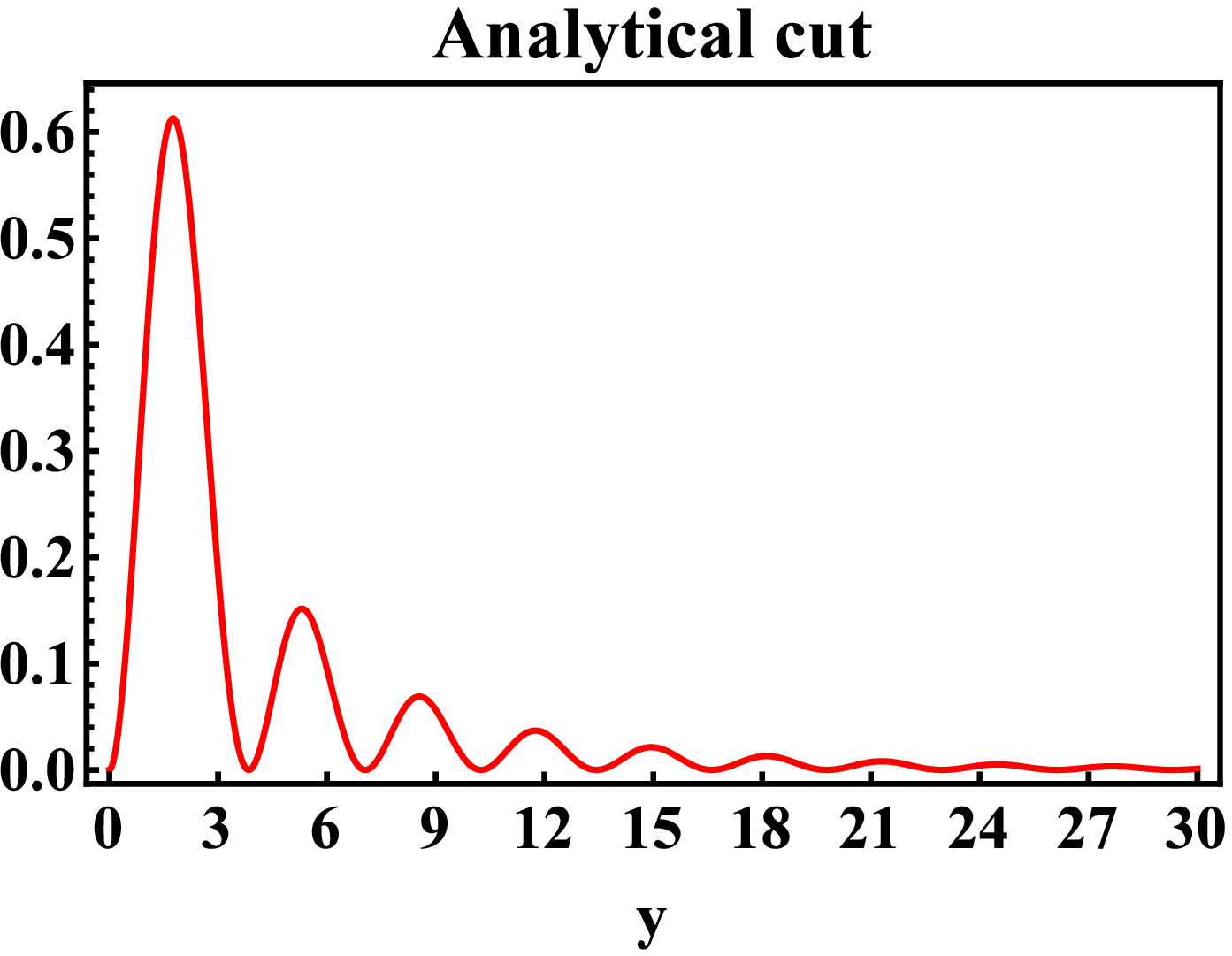} \\	 
	\end{tabular}
	\caption{
	$S^x_{\bar{2}}$ for $x$-impurity (in arbitrary units) in coordinate space, for an in-plane $\dd$ vector, and for an energy $E = E_{\bar 2}$. 
	In the left two panels we show the numerical simulations on a square lattice with spectrum $\Xi_k = \mu - 2t (\cos k_x + \cos k_y )$, where we take $\mu=3,\,t=1$ and the lattice constant is set to unity (for more details see [\onlinecite{Kaladzhyan2015}]).
	In the right two panels we plot  analytical results derived in this manuscript. To match the spectrum on a lattice we take $\nu=1/4\pi, v_F = 2, k_F = 1$. The triplet pairing $\tilde\varkappa = 0.2$ and the impurity strength $J_x = 2$ for both panels.
	The two lower panels correspond to  1D cuts of the upper two panels at $x=0$.
	It is clear that both plots reflect the characteristic p-wave four-fold symmetry and  qualitatively agree except at very short distance as expected.}
	\label{annum}
\end{figure}

In Fig.~\ref{annum} we present a qualitative comparison of these analytical results with the previously obtained numerical simulations on a square lattice by calculating $S^x_{\bar{2}}$ for a $x$-impurity (in arbitrary units) in coordinate space. The two approaches agree very well except at small distances from the impurity. This is expected since the analytical model is a low-energy approximation for the square-lattice model introduced in [\onlinecite{Kaladzhyan2015}], and thus it is expected to give accurate results at small energies and large distances. We also note that the  wave functions we calculated analytically are unnormalized ones. Therefore, the  overall amplitude of the results cannot be compared (and hence the different scales). Note there is a small discrepancy between the periods of the oscillations obtained using analytical and numerical tools. 
which can be traced back to the difference of the energies of the Shiba statesbetween the two models. Overall, the qualitative agreement between the numerical and analytical results is remarkably good, especially at large distances, as expected.

\vspace{.2in}
\subsubsection{ Out-of-plane $\bm{d_\perp}$}
\vspace{.2in}
Among the unconventional SCs, Sr$_2$RuO$_4$ is believed to be a p-wave superconductor with an out-of-plane $\dd$ vector\cite{Maeno2003}.
Contrary to the in-plane $\dd$ vector, such p-wave SC breaks time reversal symmetry. It is therefore interesting to analyze and compare it with the case of $\dd_\parallel$. The  retarded Green's function can be written as:
\begin{equation*}
G_0(E,\bm r) = 
	\begin{pmatrix}
		\left[E X_0(r) + X_1(r)\right]\sigma_0 & D_\perp(\bm r)  \\
		-D^*_\perp(\bm r) & \left[E X_0(r) - X_1(r)\right]\sigma_0
	\end{pmatrix},
\end{equation*}
where we used
\begin{align*}
D_\perp(\bm r) \equiv \begin{pmatrix} iX^+_2(\bm r) & 0 \\ 0 & -iX^+_2(\bm r) \end{pmatrix}.
\end{align*}
We proceed following the same scheme as for $\dd_\parallel$.

{\it Scalar impurity.}  For this type of impurity we find
\begin{align*}
\Phi_{\bar 1}(\bm r) = +U\begin{pmatrix} 0 \\ E_{\bar 1} X_0(r) + X_1(r) \\ 0 \\ i X^-_2(\bm r) \end{pmatrix}, \\  
\Phi_{\bar 2}(\bm r) = +U\begin{pmatrix} E_{\bar 2} X_0(r) + X_1(r) \\ 0 \\ -i X^-_2(\bm r) \\ 0 \end{pmatrix}, \\
\Phi_2(\bm r) = -U\begin{pmatrix} 0 \\ -i X^+_2(\bm r) \\ 0 \\ E_2 X_0(r) - X_1(r) \end{pmatrix}, \\
\Phi_1(\bm r) = -U\begin{pmatrix} i X^+_2(\bm r) \\ 0 \\ E_1 X_0(r) - X_1(r) \\ 0 \end{pmatrix}.
\end{align*}
We note that, similar to the case of $\bm{d_\parallel}$,  the Hamiltonian for a scalar impurity commutes with $M^z_\perp$ and therefore the states found above are also eigenstates of  $M^z_\perp$ , namely: $M^z_\perp \Phi_{\bar 1, \bar 2} = -\frac{1}{2} \Phi_{\bar 1, \bar 2}$ and $M^z_\perp \Phi_{1,2} = +\frac{1}{2} \Phi_{1,2}$. We thus have $S^x(\bm r)=S^y(\bm r)=0$, and:
\begin{align*}
	S^z_{\bar 1}(r) &= -\rho_{\bar 1}(r) = +U^2 X^-_2(\bm r) X^+_2(\bm r) \leqslant 0, \\
	S^z_{\bar 2}(r) &= + \rho_{\bar 2}(r) = -U^2 X^-_2(\bm r) X^+_2(\bm r) \geqslant 0, \\
	S^z_2(r) &= -\rho_2(r) = -U^2\left(E_{1,2} X_0(r) - X_1(r)\right)^2 \leqslant 0, \\
	S^z_{1}(r) &= +\rho_{1}(r) = +U^2\left(E_{1,2} X_0(r) - X_1(r)\right)^2 \geqslant 0,
\end{align*}
once more radially symmetric. Obviously, the total SP LDOS vanishes, as it adds up exactly to zero for both pairs of degenerate states.

{\it Magnetic impurity with spin $\parallel$z.} We find for the SBS eigenstates 
\begin{align*}
\Phi_{\bar 1}(\bm r) = +J_z\begin{pmatrix} i X^+_2(\bm r) \\ 0 \\ E_{\bar 1} X_0(r) - X_1(r) \\ 0 \end{pmatrix}, \\
\Phi_{\bar 2}(\bm r) = +J_z\begin{pmatrix} E_{\bar 2} X_0(r) + X_1(r) \\ 0 \\ -i X^-_2(\bm r) \\ 0 \end{pmatrix}, \\
\Phi_2(\bm r) = -J_z\begin{pmatrix} 0 \\ -i X^+_2(\bm r) \\ 0 \\ E_2 X_0(r) - X_1(r) \end{pmatrix}, \\
\Phi_1(\bm r) = -J_z\begin{pmatrix} 0 \\ E_1 X_0(r) + X_1(r) \\ 0 \\ i X^-_2(\bm r) \end{pmatrix}.
\end{align*}
Same as before $M^z_\perp \Phi_{\bar 1, 2} = +\frac{1}{2} \Phi_{\bar 1, 2}$ and $M^z_\perp \Phi_{1,\bar 2} = -\frac{1}{2} \Phi_{1, \bar 2}$. Thus $S^x(\bm r) = S^y(\bm r) = 0$, and 
\begin{align*}
	S^z_{\bar 1}(r) &= +\rho_{\bar 1}(r) = +J_z^2\left(E_{\bar 1} X_0(r)-X_1(r)\right)^2 \geqslant 0, \\
	S^z_{\bar 2}(r) &= +\rho_{\bar 2}(r) = -J_z^2 X^-_2(\bm r) X^+_2(\bm r) \geqslant 0,\\
	S^z_2(r) &= -\rho_2(r) = -J_z^2 \left(E_{2} X_0(r) - X_1(r)\right)^2 \leqslant 0, \\
	S^z_1(r) &= -\rho_1(r) = +J_z^2 X^-_2(\bm r) X^+_2(\bm r) \leqslant 0.
\end{align*}
It is easy to see using the definitions of $X_2^\pm(\bm r)$ that all the functions above have no angular dependence, and moreover do not change sign when varying $r$. Thus we infer that the spatially-averaged spin is positive for the states $\bar 1, \bar 2$ and negative for the states $1,2$ and thus the inner states have spins of the same sign for $\dd_\perp$, different from what we obtain for $\dd_\parallel$, for which the inner states have opposite signs. This may be used experimentally as one of the distinguishing features between these two choices of $\dd$ vectors.

{\it Magnetic impurity with spin $\parallel$x.}
Unlike for $\dd_\parallel$, for $\dd_\perp$  the Hamiltonian of a magnetic impurity with spin along $x$ commutes with $M^z_\perp$ and, therefore, the SBS are also eigenstates of $M^z_\perp$: $M^z_\perp \Phi_{\bar 1, 2} = +\frac{1}{2} \Phi_{\bar 1, 2}$ and $M^z_\perp \Phi_{1,\bar 2} = -\frac{1}{2} \Phi_{1, \bar 2}$. That is why we expect that the SBS preserve the rotational symmetry in this limit, and that no peculiar feature due to the p-wave four-fold symmetry can  be observed. Indeed
\begin{align*}
\Phi_{\bar 1}(\bm r) &= +J_x\begin{pmatrix} i X^+_2(\bm r) \\ -i X^+_2(\bm r) \\ E_{\bar 1} X_0(r) - X_1(r) \\ E_{\bar 1} X_0(r) - X_1(r) \end{pmatrix}, \\
\Phi_{\bar 2}(\bm r) &= +J_x\begin{pmatrix} E_{\bar 2} X_0(r) + X_1(r) \\ E_{\bar 2} X_0(r) + X_1(r) \\ -i X^-_2(\bm r) \\ i X^-_2(\bm r) \end{pmatrix}, \\
\Phi_2(\bm r) &= -J_x\begin{pmatrix} i X^+_2(\bm r) \\ i X^+_2(\bm r) \\ E_2 X_0(r) - X_1(r) \\ -\left( E_2 X_0(r) - X_1(r)\right) \end{pmatrix}, \\
\Phi_1(\bm r) &= -J_x\begin{pmatrix} E_1 X_0(r) + X_1(r) \\ -\left( E_1 X_0(r) + X_1(r) \right)\\ -i X^-_2(\bm r) \\ -i X^-_2(\bm r) \end{pmatrix},
\end{align*}
and for all the states $S^y(\bm r)=S^z(\bm r)=0$. The other components are given by
\begin{align*}
	S^x_{\bar 1}(r) &= \rho_{\bar 1}(r) = +2J_z^2\left(E_{\bar 1} X_0(r)-X_1(r)\right)^2 \geqslant 0, \\
	S^x_{\bar 2}(r) &= -\rho_{\bar 2}(r) = +2J_z^2 X^-_2(\bm r) X^+_2(\bm r) \leqslant 0,\\
	S^x_2(r) &= -\rho_2(r) = -2J_z^2 \left(E_{2} X_0(r) - X_1(r)\right)^2 \leqslant 0, \\
	S^x_1(r) &= +\rho_1(r) = -2J_z^2 X^-_2(\bm r) X^+_2(\bm r) \geqslant 0.
\end{align*}
It is easy to see that all the functions above have rotational symmetry and give the same sign for the spatially-averaged spin for the inner states.

\section{1D superconducting wires} \label{1Dsystem}

In what follows we consider a 1D superconducting wire directed along the $x$-axis and described by the Hamiltonian (\ref{H0}). As 1D systems cannot be intrinsic superconductors, the superconductivity in these systems needs to be induced via a proximity effect. Similarly to the 2D case, we first revisit the limit of purely s-wave pairing, and subsequently of purely p-wave pairing. 

\subsection{Pure s-wave superconductors} \label{1Dswave}
In the case of s-wave singlet pairing $\varkappa = 0$, and the unperturbed retarded Green's function in momentum space is given by:
$$
G_0(E,k) = -\frac{1}{\xi_k^2+\omega^2}
		\begin{pmatrix} (E+\xi_k) \sigma_0 & \Delta_s \sigma_0  \\ 
		 \Delta_s \sigma_0 & (E-\xi_k) \sigma_0
	\end{pmatrix},
$$
and therefore we have two types of integrals to compute:
\begin{align}
X_0(x) &= -\int \frac{dk}{2\pi} \frac{e^{ikx}}{\xi_k^2+\omega^2}, \\
X_1(x) &= -\int \frac{dk}{2\pi} \frac{\xi_k e^{ikx}}{\xi_k^2+\omega^2}, 
\end{align}
where $\omega^2 = \Delta_s^2 - E^2$. For $k>0$ we linearize the spectrum around the Fermi momentum, thus $k = k_F + \xi_k/v_F$, and we get:
\begin{align}
X_0(x) &= -\frac{1}{v_F} \cdot \frac{1}{\omega} \cdot \cos k_F x \cdot e^{-\omega |x|/v_F} 
\label{X0Ds}\\
X_1(x) &= \frac{1}{v_F} \cdot \sin k_F |x| \cdot e^{-\omega |x|/v_F}.
\label{X1Ds}
\end{align}
Since there are no divergences like in the 2D limit, these expressions can be used also to find the $x=0$ limit of the Green's functions. The full form of the Green's function can be written as:
\begin{align}
G_0(E,x) = 
	\begin{pmatrix}
			\left[E X_0(x)+X_1(x)\right]\sigma_0 & \Delta_s X_0(x) \sigma_0 \\ 
			 \Delta_s X_0(x) \sigma_0  & \left[E X_0(x)-X_1(x)\right]\sigma_0 \\
	\end{pmatrix},
\end{align}
with the $x=0$ limit being given by:
\begin{align}
G_0(E,x=0) = -\frac{1}{v_F} \frac{1}{\sqrt{\Delta_s^2-E^2}}
		\begin{pmatrix} 	
				E\sigma_0 & \Delta_s \sigma_0\\ 
				\Delta_s \sigma_0 & E \sigma_0
		\end{pmatrix}.
\end{align}
We note that this has a similar structure to the 2D Green's function described in Section \ref{2Dswave}, thus we expect to obtain similar results as in the 2D limit, with $x$ replacing $r$ and $1/v_F$ replacing $\pi \nu $.
Therefore, same as in 2D, a scalar impurity does not induce any SBS for a purely s-wave SC. In what concerns the magnetic impurities we consider an impurity with spin directed along the $z$ axis, and we thus have the following energies for the Shiba states
\begin{align}
	E_{1,\bar{1}} = \pm \frac{1-\alpha^2}{1+\alpha^2} \Delta_s, \; \text{where}\; \alpha =  J/v_F.
\end{align}
Moreover, by rewriting the expressions from Section \ref{2Dswave} in terms of (\ref{X0Ds},\ref{X1Ds}), we obtain the expression for the $S^z$-component of the SBS for a positive-energy state:
\begin{equation}
S^z_1(x) = -(\cos k_F x -\alpha \sin k_F |x|)^2 e^{- 2k_s |x|},
\end{equation}
where $k_F$ is the Fermi momentum and $k_s = \omega/v_F$ is the inverse superconducting decay length. To get the expressions for the negative-energy eigenstate one needs to replace $\alpha \to -\alpha$ and add an overall minus sign. This has a similar structure to the asymptotic form of the corresponding Friedel oscillations in the 2D limit, i.e. oscillations with a $k_F$ wavevector and an exponential spatial decay, with the only qualitative difference that in 2D the oscillations exhibit an additional power-law decay. 

\begin{figure}
	\includegraphics*[scale=0.4]{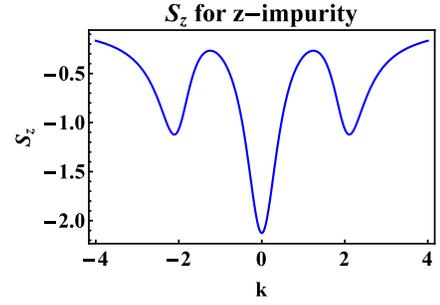} 
	\caption{The $z$-component of SP LDOS (in arbitrary units) as a function of momentum for the positive-energy Shiba state. We consider a magnetic impurity with spin along $z$ and of impurity strength $J_z = 0.25$. We take $\Delta_s=0.5$, $\varkappa = 0$ and an inverse quasiparticle lifetime of $\delta=0.01$.}
	\label{figure2}
\end{figure}

The analytical form is simple enough to perform a Fourier transform, and we obtain
\begin{eqnarray}
S^z_1(k) &= &-\frac{2(1+\alpha^2)k_s}{k^2+4k_s^2} 
-\frac{(1-\alpha^2)k_s - \alpha(k+2k_F)}{(k+2k_F)^2+4k_s^2}\nonumber\\&&
-\frac{(1-\alpha^2)k_s + \alpha(k-2k_F)}{(k-2k_F)^2+4k_s^2}.
\end{eqnarray}
We note that this expression corresponds to three high-intensity features at $k=-2k_F$, $k=0$ and $k=2k_F$, as expected given the form of the real-space oscillations with a $2k_F$ periodicity (see figure \ref{figure2}). Note also that the exponential spatial decay with a $k_s$ wavevector is translated into momentum space as a widening of the high-intensity features given exactly by $k_s$.

It is worth mentioning that the expressions for an impurity with the spin directed along the $x$ axis are exactly the same, with $S_x$ being the only non-zero component in this case.

\subsection{Pure p-wave superconductors} \label{1Dpwave}
To consider a 1D p-wave superconductor we formally choose the triplet pairing parameter to be $\dd = (0,\, -k,\, 0)$, although the concept of the $\dd$ vector is not well-defined in 1D. To find the actual gap in the spectrum we seek the minimum of the energy dispersion $\sqrt{\xi^2_k+\varkappa^2k^2}$, and thus we have
\begin{equation}
\Delta_t = \frac{\varkappa k_F}{\sqrt{1+\tilde{\varkappa}^2}},
\end{equation}
 reached at 
\begin{equation}\label{eq:kpf}
k'_F \equiv \frac{k_F}{1+\tilde{\varkappa}^2}.
\end{equation}
The retarded Green's function in momentum space can be written as:
\begin{align}
G_0(E,k) = -\frac{1}{\xi_k^2+\varkappa^2 k^2 - E^2}
		\begin{pmatrix} 	
				(E+\xi_k)\sigma_0 & -\varkappa k \sigma_y \\ 
				-\varkappa k \sigma_y & (E-\xi_k)\sigma_0 
		\end{pmatrix},
\end{align}
To obtain the form of the Shiba states we need to calculate the three following integrals:
\begin{align}
X_0(x) &= -\int \frac{dk}{2\pi} \frac{e^{ikx}}{\xi_k^2+\varkappa^2 k^2 - E^2}, \\
X_1(x) &= -\int \frac{dk}{2\pi} \frac{\xi_k \; e^{ikx}}{\xi_k^2+\varkappa^2 k^2 - E^2}, \\
X_2(x) &= -\int \frac{dk}{2\pi} \frac{i \varkappa k \; e^{ikx}}{\xi_k^2+\varkappa^2 k^2 - E^2},
\end{align}
The calculations are performed in a similar fashion as for the s-wave. For $k>0$ we linearize the spectrum around the Fermi momentum, thus $k = k_F + \xi_k/v_F$ and we get:
\begin{align}
X_0(x) &= -\frac{1}{v_F} \frac{1}{1+\tilde{\varkappa}^2} \cdot \frac{1}{\omega} \cdot \cos k'_F x \cdot e^{-\omega |x|/v_F} \\
X_1(x) &= \frac{1}{v_F} \frac{1}{1+\tilde{\varkappa}^2} \left[\frac{\gamma \Delta_t}{\omega} \cos k'_F x +\sin k'_F |x| \right] e^{-\omega |x|/v_F} \\
X_2(x) &= \frac{1}{v_F} \frac{\tilde{\varkappa}}{1+\tilde{\varkappa}^2} \times \\
&\times \left[\frac{k'_F v_F}{\omega} \sin k'_F x + \sgn x \cos k'_F x\right]  e^{-\omega |x|/v_F},
\end{align}
where $\omega^2 = \frac{\Delta_t^2 - E^2}{1+\tilde{\varkappa}^2}$. The full form of the Green's function can thus be written as:
\begin{align}
G_0(E,x) = 
	\begin{pmatrix}
			\left[ E X_0(x)+X_1(x) \right] \sigma_0 & i X_2(x) \sigma_y \\ 
			 i X_2(x) \sigma_y & \left[ E X_0(x)-X_1(x) \right] \sigma_0
	\end{pmatrix},
\end{align}
with the $x=0$ limit being given by:
\begin{align}
G_0(E,x=0) = -\frac{1}{v_F} \frac{1}{\sqrt{1+\tilde{\varkappa}^2}} \frac{1}{\sqrt{\Delta_t^2 - E^2}} \times \phantom{aaaaaaaaa}\\ \nonumber
		\times
		\begin{pmatrix} 	
				\left[ E-\gamma \Delta_t\right] \sigma_0 & 0\\ 
				0 & \left[ E+\gamma \Delta_t\right] \sigma_0
		\end{pmatrix}.
\end{align}

{\it Scalar impurity.} 
The energies of the SBS can be obtained in the same fashion as in the 2D limit, and we have
\begin{align}
E_{\bar 1, \bar 2} &= - \frac{-\gamma \beta^2 + \sqrt{1+\beta^2 (1-\gamma^2)}}{1+\beta^2}\Delta_t, \\ 
E_{1,2} &= + \frac{-\gamma \beta^2 + \sqrt{1+\beta^2 (1-\gamma^2)}}{1+\beta^2}\Delta_t,
\end{align}
where now $ \beta = \frac{U}{\sqrt{v_F^2+\varkappa^2}}$.
The SBS eigenstates are given by
\begin{align*}
\Phi_{\bar 1}(x) &= +U\begin{pmatrix} 0 \\ E_{\bar 1,\bar 2} X_0(x) + X_1(x) \\ X_2(x) \\ 0 \end{pmatrix}, \\ 
\Phi_{\bar 2}(x) &= +U\begin{pmatrix} E_{\bar 1,\bar 2} X_0(x) + X_1(x) \\ 0 \\ 0 \\ -X_2(x) \end{pmatrix}, \\
\Phi_2(x) &= -U\begin{pmatrix} X_2(x) \\ 0 \\ 0 \\ E_{1,2} X_0(x) - X_1(x) \end{pmatrix}, \\
\Phi_1(x) &= -U\begin{pmatrix} 0 \\ -X_2(x) \\ E_{1,2} X_0(x) - X_1(x) \\ 0 \end{pmatrix}.
\end{align*}
For the same reason as in the 2D case, $S_x(x)=S_y(x)=0$, whereas
\begin{align*}
	S^z_{\bar 1}(x) &= +\rho_{\bar 1}(x) = +U^2 X^2_2(x) \geqslant 0,\\
	S^z_{\bar 2}(x) &= -\rho_{\bar 2}(x) = -U^2 X^2_2(x) \leqslant 0, \\
	S^z_{2}(x) &= -\rho_{2}(x) = -U^2 \left(E_{1,2} X_0(x) - X_1(x)\right)^2 \leqslant 0, \\
	S^z_1(x) &= +\rho_1(x) = +U^2 \left(E_{1,2} X_0(x) - X_1(x)\right)^2 \geqslant 0.
\end{align*}
Similarly to the 2D systems, the spins sum up to zero for each pair of degenerate energy levels, and therefore the SP LDOS vanishes, with only the non-polarized LDOS being non-zero.

\begin{figure}
	\includegraphics*[scale=0.4]{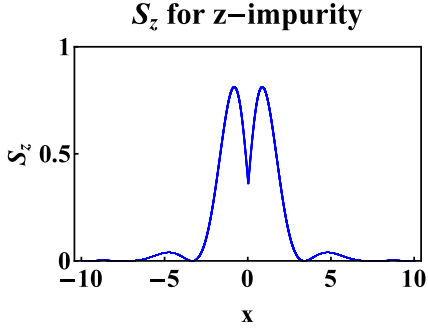} 
	\caption{The $z$-component of SP LDOS (in arbitrary units) as a function of position for an energy $E = E_{1}$. We consider a magnetic impurity with spin along $z$ and of impurity strength $J_z = 1.5$. We take $\Delta_s=0$, $\varkappa = 0.5$ and an inverse quasiparticle lifetime of $\delta=0.01$.}
	\label{figure3}
\end{figure}

{\it Magnetic impurity with spin $\parallel$ z.}
The energies of the Shiba states can be obtained along the same lines as for the 2D case,
\begin{align}
E_{1,\bar 1} &= \pm \frac{\gamma \alpha^2 + \sqrt{1+\alpha^2 (1-\gamma^2)}}{1+\alpha^2}\Delta_t, \\
E_{2,\bar 2} &= \pm \frac{-\gamma \alpha^2 + \sqrt{1+\alpha^2 (1-\gamma^2)}}{1+\alpha^2}\Delta_t,
\end{align}
with $\alpha = \frac{J}{\sqrt{v_F^2+\varkappa^2}},\;$
while the coordinate dependence is given by:
\begin{align*}
\Phi_{\bar 1}(x) &= +J_z\begin{pmatrix} 0 \\ -X_2(x) \\ E_{\bar 1} X_0(x) - X_1(x) \\ 0 \end{pmatrix}\\
\Phi_{\bar 2}(x) &= +J_z\begin{pmatrix} E_{\bar 2} X_0(x) + X_1(x) \\ 0 \\ 0 \\ -X_2(x) \end{pmatrix}\\
\Phi_2(x) &= -J_z\begin{pmatrix} X_2(x) \\ 0 \\ 0 \\ E_2 X_0(x) - X_1(x) \end{pmatrix}\\
\Phi_1(x) &= -J_z\begin{pmatrix} 0 \\ E_1 X_0(x) + X_1(x) \\ X_2(x) \\ 0 \end{pmatrix}  
\end{align*}
Take notice of $S_x(x) = S_y(x) = 0$ due to the absence of symmetry breaking in those directions. The rest of the components is given by
\begin{align*}
	S^z_{\bar 1}(x) &= +\rho_{\bar 1}(x) = +J_z^2 \left(E_{\bar 1} X_0(x) - X_1(x)\right)^2 \geqslant 0, \\
	S^z_{\bar 2}(x) &= -\rho_{\bar 2}(x) = -J_z^2 X^2_2(x) \leqslant 0, \\
	S^z_2(x) &= -\rho_2(x) = -J_z^2 \left(E_2 X_0(x) - X_1(x)\right)^2 \leqslant 0, \\
	S^z_1(x) &= +\rho_1(x) = +J_z^2 X^2_2(x) \geqslant 0.
\end{align*}
All these functions are even with respect to position. In order to illustrate this, we plot in figure \ref{figure3} the coordinate dependence of the $S_z$ component for the positive-energy state with $E=E_1$. 

The form of the eigenstates allows an analytical calculation of the Fourier transforms, same as for the pure s-wave limit. Below we give the momentum space behavior of the SP LDOS components. For the first positive-energy state with $E=E_1$ we have:
\begin{eqnarray}
&&S^{z}_1(k) = \alpha^2 \gamma^2 \left\{\left(1 + \frac{v_F^2k_F'^2}{\omega^2} \right) \frac{2k_s}{k^2+4k_s^2} \right. \\
 &&+  \left(1 - \frac{v_F^2k_F'^2}{\omega^2} \right) \left[\frac{k_s}{(k+2k'_F)^2+4k_s^2} + \frac{k_s}{(k-2k'_F)^2+4k_s^2} \right] \nonumber\\
&& \left. + \frac{v_F k_F'}{\omega} \left[\frac{k+2k'_F}{(k+2k'_F)^2+4k_s^2} - \frac{k-2k'_F}{(k-2k'_F)^2+4k_s^2} \right] \right\} \nonumber
\end{eqnarray}
Where $k_s = \omega/v_F$. For the second positive-energy state with $E=E_2$ we have
\begin{eqnarray}
&&S^{z}_2(k) = -\left(1+\frac{\alpha^2}{1+\tilde{\varkappa}^2}\right) \frac{2k_s}{k^2+4k_s^2}  \\
 &&-\left(1-\frac{\alpha^2}{1+\tilde{\varkappa}^2}\right) \left[\frac{k_s}{(k+2k'_F)^2+4k_s^2} + \frac{k_s}{(k-2k'_F)^2+4k_s^2} \right]  \nonumber\\
&&-\frac{\alpha}{\sqrt{1+\tilde{\varkappa}^2}} \left[\frac{k+2k'_F}{(k+2k'_F)^2+4k_s^2} - \frac{k-2k'_F}{(k-2k'_F)^2+4k_s^2} \right] \nonumber
\end{eqnarray}
Same as in the purely s-wave limit we see that the high-intensity features appear at three momenta $k=-2k'_F$, $k=0$, $k=+2k'_F$ with $k_s$ being responsible for the widening of the Lorentzian peaks. However we note that in the p-wave limit the Fermi momentum is renormalized to $k'_F\equiv \frac{k_F}{1+\tilde{\varkappa}^2}$ due to the triplet pairing, as described above.

{\it Magnetic impurity with spin $\parallel$ x}

Same as in the 2D case the energy levels don't depend on the impurity direction and are the same as when the impurity spin is directed along $z$, whereas the coordinate dependence changes:
\begin{align*}
\Phi_{\bar 1}(x) &= J_x\begin{pmatrix} X_2(x) \\ -X_2(x) \\  E_{\bar 1} X_0(x) - X_1(x) \\ E_{\bar 1} X_0(x) - X_1(x) \end{pmatrix}\\
\Phi_{\bar 2}(x) &= J_x\begin{pmatrix} E_{\bar 2} X_0(x) + X_1(x) \\ E_{\bar 2} X_0(x) + X_1(x) \\ X_2(x) \\ -X_2(x) \end{pmatrix}\\
\Phi_2(x) &= J_x\begin{pmatrix} X_2(x) \\ X_2(x) \\  -E_2 X_0(x) + X_1(x) \\ E_2 X_0(x) - X_1(x) \end{pmatrix}\\
\Phi_1(x) &= J_x\begin{pmatrix} -E_1 X_0(x) - X_1(x) \\ E_1 X_0(x) + X_1(x) \\ X_2(x) \\ X_2(x) \end{pmatrix}
\end{align*}
Since there is no symmetry breaking along $y$-axis and $z$-axis, $S^y(x) = S^z(x) = 0$, and
\begin{align*}
	S^x_{\bar 1}(x) &= +\rho_{\bar 1}(x) = +2J_x^2 \left( E_{\bar 1} X_0(x)- X_1(x) \right)^2 \geqslant 0, \\
	S^x_{\bar 2}(x) &= -\rho_{\bar 2}(x) = -2J_x^2 X^2_2(x) \leqslant 0,\\
	S^x_2(x) &= -\rho_2(x) = -2J_x^2 \left( E_2 X_0(x) - X_1(x) \right)^2 \leqslant 0,\\
	S^x_1(x) &= +\rho_1(x) = +2J_x^2 X^2_2(x) \geqslant 0.
\end{align*}
The coordinate dependence of these functions is similar to the one depicted in figure \ref{figure3} for the case of a $z$-impurity.

Analyzing the results obtained for an impurity in a 1D  p-wave SC, we find that, unlike for  2D p-wave SCs, there is no way to identify symmetry breaking features such as the p-wave four-fold structure observed in figure \ref{figure1}, since all the components of the SP LDOS, for all types of impurities, are even. However, we want to emphasize that the triplet-pairing parameter can still be extracted using the decay length $k_s$ and the wave vector $k_F'$ of the Friedel oscillations, both of these parameters having an explicit dependence on the value of the p-wave pairing (see (\ref{eq:kpf})).


\section{Conclusions}

We have calculated analytically the spatial structure and the asymptotic expansions of the wavefunctions for the SBS in 2D and 1D superconductors with singlet or triplet pairing.  We have shown that the strong features originating from the orbital nature of the p-wave allow to distinguish not only between the singlet and triplet pairing cases, but also between different types of triplet pairing. Our results are consistent with previous numerical results for 2D systems presented in [\onlinecite{Kaladzhyan2015}]. We believe that our results can be used for studying the topological phases of matter that can be engineered with impurities in different types of p-wave superconductors, in particular the exact analytical form of the Shiba wavefunctions, are useful for computing the Chern numbers in such emergent topological superconductors. Moreover, we propose to extract the triplet pairing parameter using two characteristic lengths -- the superconducting decay length scale and the period of Friedel oscillations. While the triplet pairing parameter can be more generally obtained as the bulk spectral gap in STM measurements, extracting the triplet pairing parameter using the Friedel oscillations may serve as an independent alternative method to consistently measure the triplet pairing parameter. These quantities should be more accessible in 1D and 2D rather than in 3D due to a weaker power-law decay ($r^{-1}$ in 2D and $r^0$ in 1D versus $r^{-2}$ in 3D). We propose to measure these characteristic lengths, as well as to test the spatial structure of the SBS using spin-polarized STM.



\section*{Acknowledgements} This work is supported by the ERC Starting Independent Researcher Grant NANOGRAPHENE 256965.
PS would like to acknowledge financial support from the  French Agence Nationale de la Recherche through the contract ANR Mistral.

\begin{widetext}
\appendix

\section{Calculation of integrals}
In this appendix we calculate the integrals characterizing the retarded Green's function coordinate dependence for the cases of pure s-wave and pure p-wave SCs in 2D.

\subsection{Integrals for pure s-wave SCs}
We linearize the spectrum $\xi_{\bm k}=v_F(k-k_F)$, we denote $\omega^2 = \Delta_s^2-E^2,\, \Omega = \omega/v_Fk_F$, and we calculate the following integrals:
\begin{align*}
X_0(\bm r) &= -\int \frac{d\bm k}{(2\pi)^2} \frac{e^{i \bm{k r}}}{\xi_{\bm k}^2 +\omega^2}, \\
X_1(\bm r) &= -\int \frac{d\bm k}{(2\pi)^2} \frac{\xi_{\bm k}\, e^{i \bm{k r}}}{\xi_{\bm k}^2+\omega^2}.
\end{align*}
To perform the integrations we use the integral representations of Bessel functions, namely:
\begin{align*}
J_0(x) &= \frac{2}{\pi} \int\limits_1^{+\infty}\frac{\sin x u}{\sqrt{u^2-1}}du \;\;\text{for}\;\; x>0, \\
K_0(-iz) &= \int\limits_1^{+\infty} \frac{e^{i u z}}{\sqrt{u^2-1}}, \;\;\text{for}\;\; \Im z > 0,
\end{align*}
where $J_0$ and $K_0$ denote the Bessel function of the first kind and the modified Bessel function of the second kind respectively. Thus we proceed:
\begin{align*}
X_0(r) &= -\nu \int d\xi_{\bm k} \int \frac{d\varphi_{\bm k}}{2\pi} \frac{e^{i (k_F+\frac{\xi_{\bm k}}{v_F})r \cos (\varphi_{\bm k} - \varphi_{\bm r})}}{\xi_{\bm k}^2 + \omega^2} = -\nu \int d\xi_{\bm k} \frac{J_0\left[\left(1+\frac{\xi_{\bm k}}{v_Fk_F}\right) k_F r \right]}{\xi_{\bm k}^2 + \omega^2} =  -\frac{\nu}{v_Fk_F} \int dW \frac{J_0\left(W k_F r \right)}{(W-1)^2 + \Omega^2} = \\
&= -\frac{\nu}{v_Fk_F} \frac{2}{\pi} \negthickspace  \int\limits_1^{+\infty} \negthickspace \frac{dU}{\sqrt{U^2-1}} \int \negthickspace dW \frac{\sin\left(k_F r U W \right)}{(W-1)^2 + \Omega^2} = 
 -\frac{\nu}{v_Fk_F} \frac{2}{\pi} \Im \negthickspace  \int\limits_1^{+\infty} \negthickspace \frac{dU}{\sqrt{U^2-1}} \int \negthickspace dW \frac{e^{i k_F r U W}}{(W-1)^2 + \Omega^2} = \\
&= -2\nu \cdot \frac{1}{\omega} \Im \negthickspace  \int\limits_1^{+\infty} \negthickspace dU\frac{e^{i(1+i\Omega)k_Fr U}}{\sqrt{U^2-1}} = -2\nu \cdot \frac{1}{\omega} \cdot \Im K_0 \left[ -i (1+i\Omega) k_F r \right], \\
X_1(r) &= -\nu \int \xi_{\bm k} d\xi_{\bm k} \int \frac{d\varphi_{\bm k}}{2\pi} \frac{e^{i (k_F+\frac{\xi_{\bm k}}{v_F})r \cos (\varphi_{\bm k} - \varphi_{\bm r})}}{\xi_{\bm k}^2 + \omega^2} = -\nu \negthickspace\int\negthickspace d\xi_{\bm k} \frac{\xi_{\bm k}\,J_0\left[\left(1+\frac{\xi_{\bm k}}{v_Fk_F}\right) k_F r \right]}{\xi_{\bm k}^2 + \omega^2} = -\nu \negthickspace\int\negthickspace dW \frac{\left(W-1\right)J_0\left(W k_F r \right)}{(W-1)^2 + \Omega^2} = \\
&= -\nu \frac{2}{\pi} \Im \negthickspace  \int\limits_1^{+\infty} \negthickspace \frac{dU}{\sqrt{U^2-1}} \int \negthickspace dW \frac{\left(W-1\right)e^{i k_F r U W}}{(W-1)^2 + \Omega^2} = -2\nu \cdot \Im \left\{ i \negthickspace  \int\limits_1^{+\infty} \negthickspace dU\frac{e^{i(1+i\Omega)k_Fr U}}{\sqrt{U^2-1}} \right\} = -2\nu \cdot \Re K_0 \left[ -i (1+i\Omega) k_F r \right]. 
\end{align*}
\subsection{Integrals for pure p-wave SCs}
We linearize the spectrum $\xi_{\bm k}=v_F(k-k_F)$, denoting
$$
\tilde{\varkappa} = \frac{\varkappa}{v_F},\; \gamma = \frac{\tilde{\varkappa}}{\sqrt{1+\tilde{\varkappa}^2}},\; \Delta_t = \frac{\varkappa k_F}{\sqrt{1+\tilde{\varkappa}^2}},\; \omega^2 = \frac{\Delta_t^2-E^2}{1+\tilde{\varkappa}^2},\; \Omega = \frac{\omega}{v_F k_F},
$$
and calculate the integrals:
\begin{align*}
X_0(\bm r) &= -\int \frac{d\bm k}{(2\pi)^2} \frac{e^{i \bm{k r}}}{\xi_{\bm k}^2+\varkappa^2 \bm{k}^2 - E^2}, \\
X_1(\bm r) &= -\int \frac{d\bm k}{(2\pi)^2} \frac{\xi_{\bm k}\, e^{i \bm{k r}}}{\xi_{\bm k}^2+\varkappa^2 \bm{k}^2 - E^2}, \\
X^\pm_2(\bm r) &= \pm \negthickspace\int \frac{d\bm k}{(2\pi)^2} \frac{i \varkappa k_\pm\, e^{i \bm{k r}}}{\xi_{\bm k}^2+\varkappa^2 \bm{k}^2 - E^2},
\end{align*}
\begin{align*}
X_0(r) &= -\frac{\nu}{1+\tilde{\varkappa}^2}\int d\xi_{\bm k} \int \frac{d\varphi_{\bm k}}{2\pi} \frac{e^{i (k_F+\frac{\xi_{\bm k}}{v_F})r \cos (\varphi_{\bm k} - \varphi_{\bm r})}}{(\xi_{\bm k}+\gamma \Delta_t)^2 + \omega^2} = -\frac{\nu}{1+\tilde{\varkappa}^2}\int d\xi_{\bm k} \frac{J_0\left[\left(1+\frac{\xi_{\bm k}}{v_Fk_F}\right) k_F r \right]}{(\xi_{\bm k}+\gamma \Delta_t)^2 + \omega^2} = \\
&=  -\frac{\nu}{1+\tilde{\varkappa}^2} \frac{1}{v_F k_F} \int dW \frac{J_0\left(W k_F r \right)}{(W+\gamma^2-1)^2 + \Omega^2} =  -\frac{\nu}{1+\tilde{\varkappa}^2} \frac{1}{v_F k_F} \frac{2}{\pi} \negthickspace  \int\limits_1^{+\infty} \negthickspace \frac{dU}{\sqrt{U^2-1}} \int \negthickspace dW \frac{\sin\left(k_F r U W \right)}{(W+\gamma^2-1)^2 + \Omega^2} = \\
&= -\frac{\nu}{1+\tilde{\varkappa}^2} \frac{1}{v_F k_F} \frac{2}{\pi} \Im \negthickspace \int\limits_1^{+\infty} \negthickspace \frac{dU}{\sqrt{U^2-1}} \int \negthickspace dW \frac{e^{ i k_F r U W}}{(W+\gamma^2-1)^2 + \Omega^2} = -\frac{2\nu}{1+\tilde{\varkappa}^2} \frac{1}{\omega} \Im \negthickspace \int\limits_1^{+\infty} \negthickspace \frac{dU}{\sqrt{U^2-1}} e^{i k_F r \left(1-\gamma^2+i\Omega \right)U} = \\
&= -\frac{2\nu}{1+\tilde{\varkappa}^2} \cdot \frac{1}{\omega} \cdot \Im K_0 \left[ -i (1-\gamma^2+i\Omega) k_F r \right],\\
X_1(r) &= -\frac{\nu}{1+\tilde{\varkappa}^2} \negthickspace\int \negthickspace d\xi_{\bm k} \negthickspace\int \negthickspace \frac{d\varphi_{\bm k}}{2\pi} \frac{\xi_{\bm k}\,e^{i (k_F+\frac{\xi_{\bm k}}{v_F})r \cos (\varphi_{\bm k} - \varphi_{\bm r})}}{(\xi_{\bm k}+\gamma \Delta_t)^2 + \omega^2} = -\frac{\nu}{1+\tilde{\varkappa}^2}\int d\xi_{\bm k} \frac{\xi_{\bm k}\, J_0\left[\left(1+\frac{\xi_{\bm k}}{v_Fk_F}\right) k_F r \right]}{(\xi_{\bm k}+\gamma \Delta_t)^2 + \omega^2} = \\
&=  -\frac{\nu}{1+\tilde{\varkappa}^2}\int dW \frac{\left( W-1\right)J_0\left(W k_F r \right)}{(W+\gamma^2-1)^2 + \Omega^2} = -\frac{\nu}{1+\tilde{\varkappa}^2} \frac{2}{\pi} \Im \negthickspace \int\limits_1^{+\infty} \negthickspace \frac{dU}{\sqrt{U^2-1}} \int \negthickspace dW \frac{\left( W-1\right)e^{ i k_F r U W}}{(W+\gamma^2-1)^2 + \Omega^2} = \\
&= -\frac{2\nu}{1+\tilde{\varkappa}^2} \Im \negthickspace \int\limits_1^{+\infty} \negthickspace \frac{dU}{\sqrt{U^2-1}} \left(i -\frac{\gamma^2}{\Omega} \right) e^{i k_F r \left(1-\gamma^2+i\Omega \right)U} = -\frac{2\nu}{1+\tilde{\varkappa}^2} \cdot \Im \left\{ \left(i -\frac{\gamma^2}{\Omega} \right) K_0 \left[ -i (1-\gamma^2+i\Omega) k_F r \right] \right\}, \\
X_2^\pm(\bm r) &= \pm \frac{i\varkappa\nu}{1+\tilde{\varkappa}^2}\negthickspace\int \negthickspace k d\xi_{\bm k} \negthickspace \int \negthickspace \frac{d\varphi_{\bm k}}{2\pi} \frac{e^{\pm i \varphi_{\bm k}}e^{i (k_F+\frac{\xi_{\bm k}}{v_F})r \cos (\varphi_{\bm k} - \varphi_{\bm r})}}{(\xi_{\bm k}+\gamma \Delta_t)^2 + \omega^2} = \mp \frac{\varkappa k_F \cdot \nu}{1+\tilde{\varkappa}^2} \cdot e^{\pm i \varphi_{\bm r}} \negthickspace\int \negthickspace d\xi_{\bm k} \frac{\left(1+\frac{\xi_{\bm k}}{v_Fk_F}\right)\negthickspace J_1 \negthickspace \left[ \left(1+\frac{\xi_{\bm k}}{v_Fk_F}\right) k_F r\right]}{(\xi_{\bm k}+\gamma \Delta_t)^2 + \omega^2} = \\
&=\pm \frac{\varkappa k_F \cdot \nu}{1+\tilde{\varkappa}^2} \cdot e^{\pm i \varphi_{\bm r}} \cdot \frac{\partial}{\partial\left( k_F r \right)}\int \negthickspace d\xi_{\bm k} \frac{J_0 \left[ \left(1+\frac{\xi_{\bm k}}{v_Fk_F}\right) k_F r\right]}{(\xi_{\bm k}+\gamma \Delta_t)^2 + \omega^2} = \\
& = \pm \frac{2\nu}{1+\tilde{\varkappa}^2} \cdot \frac{\varkappa k_F}{\omega} e^{\pm i \varphi_{\bm r}} \times \Re \left\{\left(1-\gamma^2+i\Omega\right) K_1 \left[ -i (1-\gamma^2+i\Omega) k_F r \right]\right\},
\end{align*}
where 
$$
e^{\pm i \varphi_{\bm r}} = \frac{x \pm i y}{\sqrt{x^2+y^2}} = \frac{x \pm i y}{r}.
$$
Note that the integrals calculated for the pure p-wave case must coincide with the ones for s-wave provided $\Delta_t \to \Delta_s$ and $\varkappa = 0$ (and thus $\gamma = 0$). As expected this substitution shows that the results of the integrations are consistent. \\

\section{Asymptotic expansions}
In this appendix we give the asymptotic expansions for the modified Bessel functions of the second kind $K_0$ and $K_1$. It is known that:
\begin{align*}
K_\nu(z) \sim \sqrt{\frac{\pi}{2}} \frac{e^{-z}}{\sqrt{z}} \left[1+\mathrm{O}\left(\frac{1}{z}\right) \right] \;\;\text{for}\; |z| \to \infty
\end{align*}
Since the asymptotic form is independent of $\nu$, we will omit it below. For the case of pure s-wave:
\begin{align*}
K\!\left[ -i (1+i\Omega) k_F r \right] \!\sim\! \frac{e^{i(1+i\Omega) k_F r}}{\sqrt{-i(1+i\Omega) k_F r}} \!=\! \frac{e^{i k_F r}}{\sqrt{\Omega-i}} \frac{e^{-k_s r}}{\sqrt{k_F r}} = *
\end{align*}

\begin{align*}
\sqrt{\Omega-i} = \left(1+\Omega^2\right)^{1/4} e^{-i\theta/2}, \; \text{where}\; \theta = \arctan\frac{1}{\Omega}
\end{align*}
Therefore
\begin{align*}
*=\frac{1}{\left(1+\Omega^2\right)^{1/4}} \frac{e^{i(k_F r + \theta/2)}}{\sqrt{k_F r}} e^{-k_S r} \approx  \frac{e^{i(k_F r + \pi/4)}}{\sqrt{k_F r}} e^{-k_s r},
\end{align*}
where $k_S = \omega/v_F$. The approximation is valid since for all subgap energies since $\Omega \ll 1$. Thus
\begin{align*}
\Re K\!\left[ -i (1+i\Omega) k_F r \right] \sim \frac{\cos \left(k_F r + \pi/4 \right)}{\sqrt{k_F r}} e^{-k_s r}, \\
\Im K\!\left[ -i (1+i\Omega) k_F r \right] \sim \frac{\sin \left(k_F r + \pi/4 \right)}{\sqrt{k_F r}} e^{-k_s r}. 
\end{align*}
Similarly, for the case of pure p-wave we get:
\begin{align*}
\Re K\!\left[ -i (1-\gamma^2+i\Omega) k_F r \right] \sim \frac{\cos \left(k'_F r + \pi/4 \right)}{\sqrt{k'_F r}} e^{-k_s r}, \\
\Im K\!\left[ -i (1-\gamma^2+i\Omega) k_F r \right] \sim \frac{\sin \left(k'_F r + \pi/4 \right)}{\sqrt{k'_F r}} e^{-k_s r} ,
\end{align*}
where $k'_F = k_F (1-\gamma^2) = \frac{k_F}{1+\tilde{\varkappa}^2}$ and $k_s = \omega/v_F$.\\

\end{widetext}

\bibliography{biblio_A}

\end{document}